\documentclass[twocolumn,tighten,helvetica]{aastex631}

\def\kms{\hbox{km$\;$s$^{-1}$}}
\def\hbeta{H$\mathrm{\beta}$}
\def\halpha{H$\mathrm{\alpha}$}
\def\Fe{\hbox{\ion{Fe}{i}~617.3}}
\def\xt{$x$-$t$}
\def\SST{Swedish 1-m Solar Telescope} 
\DeclareRobustCommand{\ion}[2]{\textup{#1\,\textsc{\lowercase{#2}}}}
\usepackage{mwe}

\usepackage{hyperref}

\hypersetup{linkcolor=blue,citecolor=blue,filecolor=cyan,urlcolor=blue}

\shorttitle{Chromosphere underneath a CBP}
\shortauthors{Bose et al.}
\graphicspath{{./}{figures/}}
\begin{document}

\title{The chromosphere underneath a Coronal Bright Point}

\author[0000-0002-2180-1013]{Souvik Bose}
\affiliation{Lockheed Martin Solar \& Astrophysics Laboratory, Palo Alto, CA 94304, USA}
\affiliation{Bay Area Environmental Research Institute, NASA Research Park, Moffett Field, CA 94035, USA}
\affiliation{Institute of Theoretical Astrophysics, University of Oslo, PO Box 1029, Blindern 0315, Oslo, Norway}
\affiliation{Rosseland Centre for Solar Physics, University of Oslo, PO Box 1029, Blindern 0315, Oslo, Norway}
\email{bose@baeri.org}

\author[0000-0002-7788-6482]{Daniel N\'obrega-Siverio}
\affiliation{Instituto de Astrof\'isica de Canarias,    E-38205 La Laguna,  Tenerife, Spain}
\affiliation{Universidad de La Laguna, Dept. Astrof\'isica, E-38206 La  Laguna, Tenerife, Spain}
\affiliation{Institute of Theoretical Astrophysics, University of Oslo, PO Box 1029, Blindern 0315, Oslo, Norway}
\affiliation{Rosseland Centre for Solar Physics, University of Oslo, PO Box 1029, Blindern 0315, Oslo, Norway}

\author[0000-0002-8370-952X]{Bart De Pontieu}
\affiliation{Lockheed Martin Solar \& Astrophysics Laboratory, Palo Alto, CA 94304, USA}
\affiliation{Institute of Theoretical Astrophysics, University of Oslo, PO Box 1029, Blindern 0315, Oslo, Norway}
\affiliation{Rosseland Centre for Solar Physics, University of Oslo, PO Box 1029, Blindern 0315, Oslo, Norway}

\author[0000-0003-2088-028X]{Luc Rouppe van der Voort}
\affiliation{Institute of Theoretical Astrophysics, University of Oslo, PO Box 1029, Blindern 0315, Oslo, Norway}
\affiliation{Rosseland Centre for Solar Physics, University of Oslo, PO Box 1029, Blindern 0315, Oslo, Norway}
% \author{Butler Burton}
% \affiliation{Leiden University}
% \affiliation{AAS Journals Associate Editor-in-Chief}

% \author{Amy Hendrickson}
% \altaffiliation{AASTeX v6+ programmer}
% \affiliation{TeXnology Inc.}

% \author{Julie Steffen}
% \affiliation{AAS Director of Publishing}
% \affiliation{American Astronomical Society \\
% 1667 K Street NW, Suite 800 \\
% Washington, DC 20006, USA}

% \author{Magaret Donnelly}
% \affiliation{IOP Publishing, Washington, DC 20005}



\begin{abstract}

Coronal Bright Points (CBPs) are sets of small-scale coronal loops, connecting opposite magnetic polarities,  primarily characterized by their enhanced extreme-ultraviolet (EUV) and X-ray emission. Being ubiquitous, they are thought to play an important role in heating the solar corona. We aim at characterizing the barely-explored chromosphere underneath CBPs, focusing on the related spicular activity and on the effects of small-scale magnetic flux emergence on CBPs. We used high-resolution observations of a CBP in \hbeta{} and Fe I 617.3 nm from the \SST{} (SST) in coordination with the Solar Dynamics Observatory (SDO).  This work presents the first high-resolution observation of spicules imaged in \hbeta{}. The spicules were automatically detected using advanced image processing techniques, which were applied to the Dopplergrams derived from \hbeta{}. Here we report their abundant occurrence close to the CBP ``footpoints", and find that the orientation of such spicules is aligned along the EUV loops, indicating that they constitute a fundamental part of the whole CBP magnetic structure. Spatio-temporal analysis across multiple channels indicates that there are coronal propagating disturbances associated with the studied spicules, producing transient EUV intensity variations of the individual CBP loops. Two small-scale flux emergence episodes appearing below the CBP were analyzed; one of them leading to quiet-sun Ellerman bombs and enhancing the nearby spicular activity. This paper presents unique evidence of the tight coupling between the lower and upper atmosphere of a CBP, thus helping to unravel the dynamic phenomena underneath CBPs and their impact on the latter.

%248 words
\end{abstract}


\section{Introduction} \label{sec:intro}

Coronal Bright Points (CBPs) appear as bright, enhanced, blob-like structures when observed in the extreme-ultraviolet (EUV) light or X-rays. First observed in X-rays with the grazing incidence X-ray telescope sounding rocket mission \citep{1973SoPh...32...81V}, CBPs comprise small-scale magnetic loops connecting opposite polarities where the confined plasma is heated up to a million degrees presumably by magnetic reconnection \citep[see][]{1994ApJ...427..459P}. CBPs are ubiquitously observed in the coronal holes, quiet-Sun, and in the close vicinity of active regions alike, which makes them interesting from the perspective of their role in coronal heating. Their lifetimes range from a few hours to even a few days \citep{1974ApJ...189L..93G,2005SoPh..228..285M} and, depending upon the wavelength of observation, they appear as roundish blobs with diameters ranging between 5--30\arcsec on average \citep{1973SoPh...32...81V,1990ApJ...352..333H,2018A&A...619A..55M}. Different studies based on emission spectroscopy and imaging \citep[as discussed in the recent review by][]{2019LRSP...16....2M} suggest that the heights over which CBPs extend in the corona ranges between 5--10~Mm above the photosphere with an average of 6.5~Mm during their lifetime.

Though CBPs have been the subject of intensive research ever since their discovery back in the early 1970s (\citealp{2019LRSP...16....2M}), there are still fundamental open questions regarding these ubiquitous phenomena. For instance, 
the CBP chromospheric counterpart remains largely unexplored to date,  which may be attributed to the lack of adequate observations that target the corona and chromosphere simultaneously. To the best of our knowledge,  only two observational studies -- \citet{1981SoPh...69...77H} and \citet{2021A&A...646A.107M} have focused on this particular atmospheric layer, both finding that strong intensity enhancements in the corona preceded lower temperature (chromospheric and transition region (TR)) enhancements, thereby indicating a scenario where the heating takes place first in the corona and is later conducted toward the TR via thermal conduction. Another open question is related to the role of magnetic flux emergence on CBPs. For example, 
magnetic flux emergence is not only known to be responsible for the origin of nearly half of the CBPs \citep{2018A&A...619A..55M}, but also to enhance the chromospheric activity and associated coronal emission \citep{2021A&A...646A.107M}. So far in the CBP literature, the focus has primarily been on large-scale emergence episodes that last for several tens of minutes to hours, therefore studies about the impact of small-scale magnetic flux emergence episodes are scarce: the lack of high-resolution, coordinated magnetograms seems to be a major impediment in this regard.

The aim of this paper is to better understand the chromospheric scenery underneath a CBP with a focus on spicules and the atmospheric responses to small-scale flux emergence episodes. Spicules are one of the most abundant and ubiquitous features observed in the solar chromosphere. They are highly dynamic, thin, (multi)threaded, and elongated structures that permeate both the active and non-active regions alike \citep{2012ApJ...759...18P}.
%
%Their research has been of great importance in solar physics \citep[see extended reviews by][for example]{1972ARA&A..10...73B,2000SoPh..196...79S} ever since their discovery in 1871. 
They are broadly divided into two categories--type I and II, with the latter being more dynamic, with higher apparent velocities, shorter lifetimes, and undergoing vigorous swaying and torsional motion \citep{2007PASJ...59S.655D,2012ApJ...759...18P,2021A&A...647A.147B}. The signatures of type-II spicules are often found in the TR and coronal passbands which makes their studies exciting from the perspective of heating and mass-loading of the solar corona \citep{2009ApJ...701L...1D,2011Sci...331...55D,2014ApJ...792L..15P,2015ApJ...799L...3R,2016ApJ...820..124H,2019Sci...366..890S}. The on-disk counterparts of type-II spicules, termed as rapid blue-shifted and red-shifted excursions \citep[RBEs and RREs, see][]{2009ApJ...705..272R,2012ApJ...752..108S,2019A&A...631L...5B}, abundantly occur in the close vicinity of strong magnetic field regions \citep[such as bi-polar/unipolar field patches, see][for example]{2012ApJ...752..108S,2021A&A...647A.147B}. This makes their study also interesting in the context of CBPs since their loops appear to be rooted to strong bi-polar magnetic field configurations present in the photosphere. Multi-dimensional numerical models, e.g. by \citet{2018ApJ...864..165W} and more recently by \citet{2022ApJ...935L..21N} suggest that the loops associated with CBPs may have some relationship with jets or spicules observed deeper in solar the atmosphere, which may contribute toward transient intensity variations in the CBPs.
%Consequently, we attempt to explore whether such spicules have a relationship with the dynamics associated with CBPs and if they are a part of the same magnetic structure. 
% Therefore, we attempt to explore whether such spicules have a relationship with the dynamics associated with the CBP and are a part of the same magnetic structure.
Regarding small-scale magnetic flux emergence, our attempt to explore its effects on already existing CBPs is motivated by two very recent papers: \cite{2022ApJ...929..103T}, which find tiny EUV bright dot-like substructures inside a CBP that seem to be associated with small flux emergence episodes; and \cite{2022ApJ...935L..21N}, which argue that flux emergence occurring in a few granules may be enough to destabilize a CBP and lead to eruptions.

To achieve our objectives, we use a high-quality, ground-based dataset from the Swedish 1-m Solar Telescope \citep[SST,][]{2003SPIE.4853..341S} in coordination with the Atmospheric Imaging Assembly \citep[AIA,][]{2012SoPh..275...17L} instrument on-board NASA's Solar Dynamics Observatory \citep[SDO,][]{2012SoPh..275....3P}. For the first time, we employ high-resolution images of the chromospheric \hbeta{} spectral line to study the spicule-CBP relationship. Moreover, the impact of multiple small-scale photospheric flux emergence episodes on the chromospheric and coronal activity are also investigated from coordinated, high-resolution magnetic field measurements.

The rest of the paper is divided as follows. Section~\ref{section:observation} describes the observations and standard data reduction processes. Section~\ref{section:methods} details the methodology employed to detect on-disk spicules from SST observations and enhancing the AIA images. We show the results and discuss their significance in Sect.~\ref{section:results_discussions}, before finally summarizing and concluding the paper in Sect.~\ref{section:conclusions}.

\section{Observations and data reduction}
\label{section:observation}

\subsection{Swedish 1-m Solar Telescope}
\label{subsection:SST_obs}
\begin{figure*}[ht!]
% \plotone{Context_CBP_v2.png}
\includegraphics[width=\linewidth]{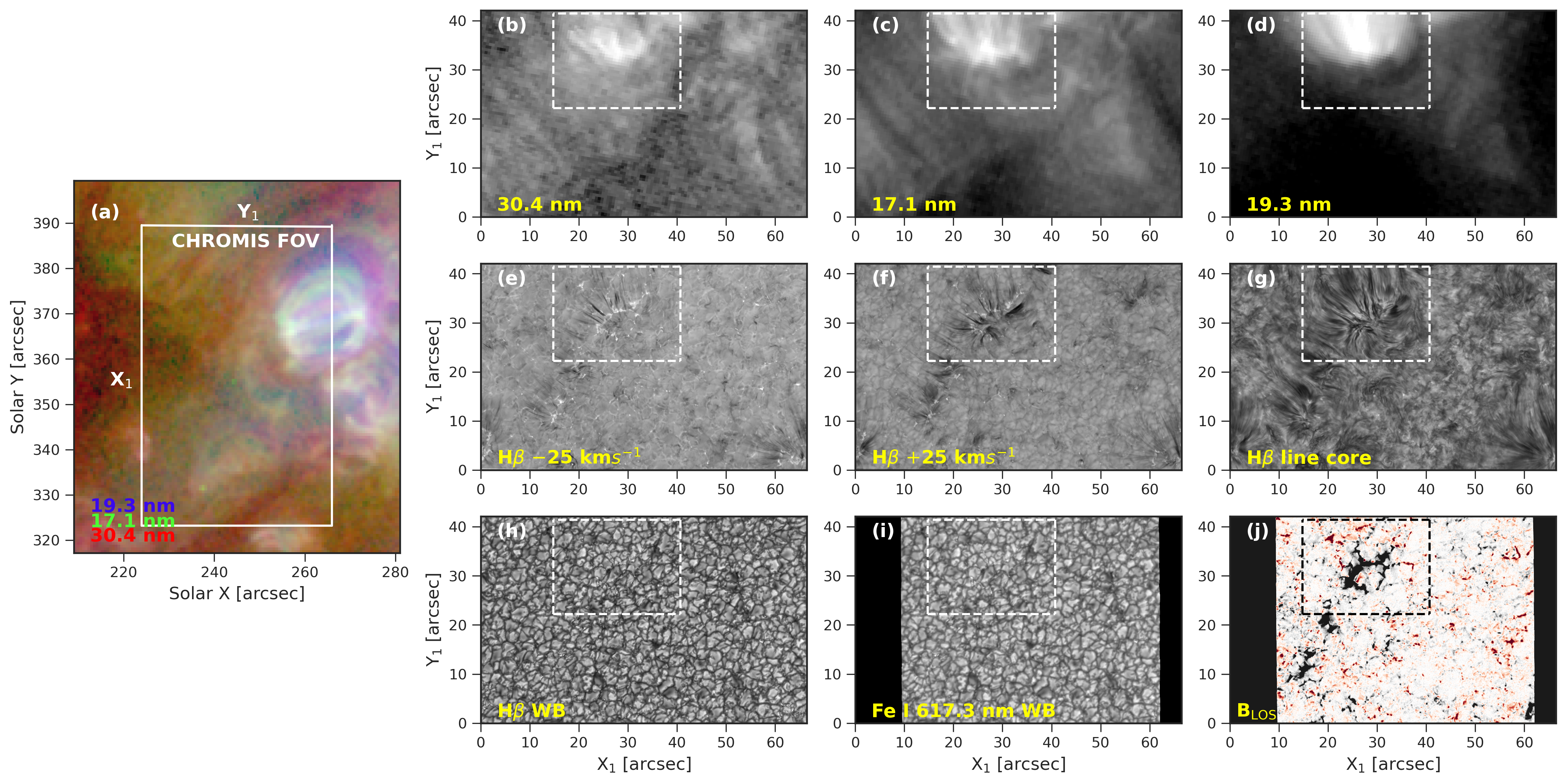}
\caption{Overview of the targeted CBP observed on 4 August 2021 at 10:03:31UT. Panel~(a) shows an RGB composite image of the CBP and its neighboring area at the original SDO/AIA pixel scale. Red, blue, and green colors correspond to 30.4, 19.3 and 17.1~nm channels, respectively. The SST/CHROMIS pointing and FOV is overlaid as a reference. Panels~(b)--(d) illustrate SDO/AIA 30.4, 17.1 and 19.3~nm channels that are rotated and co-aligned to CHROMIS. Panels~(e)--(g) show CHROMIS \hbeta{} images at blue wing ($-$~25~\kms{}), red wing ($+$~25~\kms{}) and line center, respectively. These images depict the chromospheric scene underneath the CBP. Panels~(h) and (i) contain the photospheric \hbeta{} and \ion{Fe}{i}~617.3~nm WB images, and panel~(j) shows the photospheric LOS magnetic field map (B$_{\mathrm{LOS}}$) saturated between $\pm$~60~G (black indicated positive polarity). The dashed FOV shown in panels~(a)--(j) denotes the region-of-interest associated with the CBP which forms the basis for all investigations carried out in this paper. \label{fig:context}}
\end{figure*}

For the purpose of this study, we recorded the chromospheric counterparts of the CBP using observations from the CHROMospheric Imaging Spectrometer \citep[CHROMIS,][]{2017psio.confE..85S} and CRisp Imaging Spectropolarimeter \citep[CRISP,][]{Crisp_2008} instruments at the SST on 4 August 2021, under excellent seeing conditions. The coordinates of the target were centered around solar ($X$,$Y$) = (250\arcsec,358\arcsec) with $\mu=\cos \theta=0.88$ ($\theta$ being the heliocentric angle), and the observation sequence lasted for about 11 min starting at 09:56~UTC. Figure~\ref{fig:context} shows an overview of the observed target.

CHROMIS sampled the \hbeta{} spectral line centered at 486.1~nm under imaging spectroscopic mode across 27 wavelength points between $\pm$~0.21~nm with respect to the line center. The sampling was uniform between $\pm$~0.1~nm with 0.01~nm steps. Beyond this a non-uniform sampling was intentionally chosen so as to avoid the effect of blends. Panels~(e)--(g) of Fig.~\ref{fig:context} show the \hbeta{} blue (at a Doppler offset of $-$25~\kms{}), red wing (at a Doppler offset of $+$25~\kms{}), and line core images, respectively. The cadence of the data was 6.8~s with a spatial sampling of 0\farcs038. CHROMIS also recorded wideband (WB) images with the help of an auxillary WB channel centered at 484.5~nm (referred to as \hbeta{} WB in panel~h). Besides providing context photospheric images, the WB serves as an anchor channel that aids in image restoration. The WB images have the same cadence as the narrowband \hbeta{} sequence. 

CRISP sampled the \Fe{}~nm line across 14 wavelength points under imaging spectropolarimetric mode between $-$0.032~nm and $+$0.068~nm with respect to the line center. The full Stokes \Fe{}~nm data were inverted using by using a parallel C++/Python implementation\footnote{\url{https://github.com/jaimedelacruz/pyMilne}} of the Milne-Eddington (ME) inversion scheme developed by \citet{2019A&A...631A.153D} to infer the photospheric vector magnetic field information. In addition, the \ion{Ca}{ii}~854.2~nm line was sampled across 4 wavelength points between $-$0.1~nm and $+$0.05~nm with respect to the line core in steps of 0.05~nm under imaging spectroscopic mode. The overall cadence of the combined observation sequences was measured to be 18.5~s with a spatial sampling of 0\farcs058. In this paper, we only focus on the line-of-sight (LOS) magnetic fields inferred from the \Fe{} spectral line as shown in panel~(j) of Fig.~\ref{fig:context}. 

The combination of excellent seeing conditions, the SST adaptive optics system, the high-quality CRISP and CHROMIS re-imaging systems \citep{2019A&A...626A..55S}, and Multi-Object Multi-Frame Blind Deconvolution \citep[MOMFBD,][]{2005SoPh..228..191V} image restoration resulted in high-spatial resolution data down to the diffraction limit of the telescope 
%(given by 1.22$\lambda$/$D$, where $\lambda$ is the wavelength and $D$ represents the effective diameter of the telescope). 
(for \hbeta{} 1.22$\lambda/D=0\farcs13$ with $D=0.97$~m the effective aperture of SST).
The SSTRED reduction pipeline \citep{2015A&A...573A..40D,2021A&A...653A..68L} was used to facilitate reduction of the data, including the spectral consistency technique described in \citet{2012A&A...548A.114H}. Furthermore, both the CRISP and CHROMIS time series were destretched to compensate for the residual warping across the field-of-view (FOV) which was not accounted for by the image restoration techniques described earlier.

For this study, the CRISP data (with a lower spatial and temporal resolution) were co-aligned to CHROMIS by expanding the former to CHROMIS pixel scale followed by a cross-correlation between the respective photospheric WB channels shown in panels~(h) and (i) of Fig.~\ref{fig:context}. In other words, the CHROMIS data with a FOV of $66\arcsec \times 42\arcsec$ and a cadence of 6.8~s served as a reference for the CRISP data to which the latter was aligned. We used nearest neighbour interpolation for the temporal alignment. 

\subsection{Solar Dynamics Observatory}
\label{subsection:SDO_obs}

The coronal part associated with the CBP was observed with the AIA instrument on-board SDO. The SDO datasets were co-aligned to SST (CHROMIS) datasets in the following manner. The SDO image cutout sequences were first downloaded from the Joint Science Operations Center's (JSOC) website\footnote{\url{http://jsoc.stanford.edu/}}. Next, the images from all the AIA channels were co-aligned to HMI continuum images (here the AIA 30.4~nm channel was aligned to HMI continuum), followed by the latter's co-alignment to CHROMIS WB channels via an iterative cross-correlation algorithm. Finally, the SDO images were cropped to have the same FOV as SST. The end result of this pipeline is a co-aligned SDO dataset that consists of eleven (nine AIA and two HMI) image sequences that are expanded from their original pixel scale to CHROMIS pixel scale of 0\farcs038 and matched in time by nearest-neighbour sampling to CHROMIS temporal cadence. We used the publicly available\footnote{\url{https://robrutten.nl/rridl/00-README/sdo-manual.html}} Interactive Data Language (IDL) based automated pipeline developed by Rob Rutten for this purpose \citep{2020arXiv200900376R} and refer to \citet{2021A&A...654A..51B} for an example of this pipeline's application.

% One of the major challenges in using (and further co-aligning) the SDO cutout sequences in this fashion is the internal co-alignment between the different AIA channels. This is simply because the thermal sensitivity of the different AIA channels ranges from a few tens of thousands to few million Kelvin \citep{2012SoPh..275...41B}. The pipeline attempts to circumvent this issue by downloading two sets of SDO cutout sequences -- one at full 12~s cadence centered around the SST target and the other at a lower cadence with a FOV of $700\arcsec \times 700\arcsec$ centered around the disk-center. The latter dataset is used to find the spatial offsets between the different wavelength channels by applying the height-of-formation differences in an iterative fashion. This is described further in \citet{2020arXiv200900376R}. The end result of this pipeline is a co-aligned SDO dataset that consists of eleven (nine AIA and two HMI) image sequences that are expanded from their original pixel scale to CHROMIS pixel scale of 0\farcs038 and matched in time by nearest-neighbour sampling to CHROMIS temporal cadence. 

An RGB composite image, consisting of AIA 30.4, 17.1 and 19.3~nm channels, of the CBP target at the original AIA resolution is shown in Fig.~\ref{fig:context} panel~(a), while panels~(b)--(d) show the same three channels but rotated and co-aligned to the CHROMIS data using the procedure described in this section. This co-aligned SST and SDO dataset was then visualized extensively with CRISPEX \citep{2012ApJ...750...22V}, an IDL widget-based tool that allows an efficient simultaneous exploration of multi-dimensional and multi-wavelength datasets.

\section{Methods employed}
\label{section:methods}
\subsection{Detecting on-disk spicules from \hbeta{}}
\label{subsection:spicule_detection}

\begin{figure*}[ht!]
% \plotone{Context_CBP_v2.png}
\includegraphics[width=\linewidth]{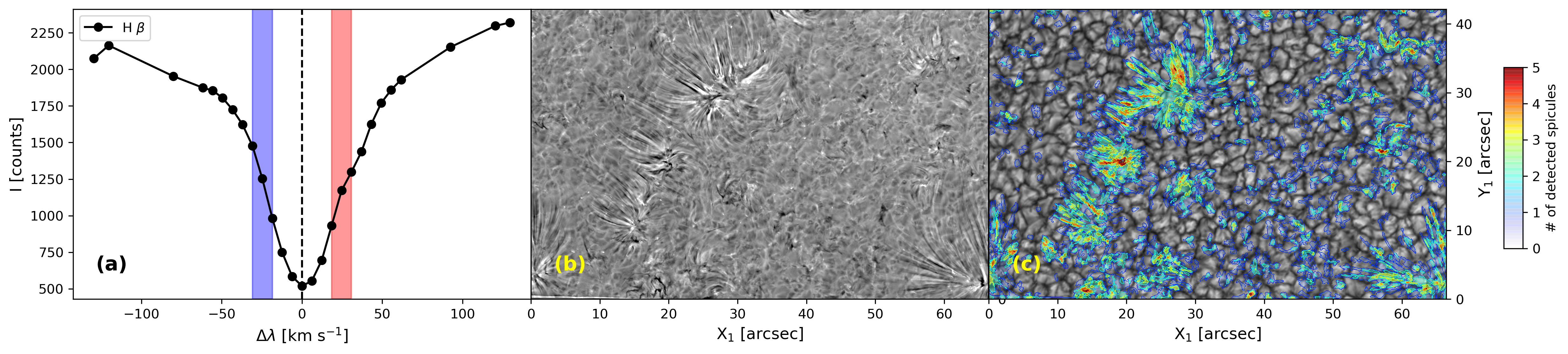}
\caption{Overview of the automated on-disk spicule detection method described in the text. Panel~(a) shows the spatio-temporal average of the \hbeta{} spectral line computed over the entire CHROMIS FOV, and further illustrates how Dopplergrams are generated by subtracting signals in the blue wing from the red wing (indicated by the shaded areas on either side of the line center). Panel~(b) shows an example of a generated Dopplergram where RBEs and RREs show up as dark and bright threaded structures. Panel~(c) shows the location and the density distribution of the detected spicules against a background of temporally averaged \hbeta{} WB image.  \label{fig:detect_spicules}}
\end{figure*}

We employed an automated detection method based on the difference between images observed in the blue and red wings of the \hbeta{} spectral line. This is similar to constructing Dopplergrams \citep[see][]{2012ApJ...752..108S,2014Sci...346D.315D,2016ApJ...824...65P}, but instead of subtracting fixed wavelengths on opposite sides of the line center, an average over a range of wavelengths (between $\pm$20--30~\kms{} on opposite sides of the line center) is computed which are then subtracted from one another as shown in Fig.~\ref{fig:detect_spicules}~(a). The difference images are then subjected to unsharp masking which causes an enhancement in the high spatial frequency components of the image. In this case, it amplifies the threaded spicular features as seen in the panel~(b). RBEs appear as darker threads with negative intensity values whereas RREs appear brighter with positive intensity values in these difference images. It is important to note that the difference maps so obtained (as in panel~b) do not correspond to absolute measure of the Doppler velocity associated with RBEs and RREs. The chief goal is to obtain a representation of the spatiotemporal evolution of the velocity patterns associated with these features.

Next, an adaptive intensity thresholding technique was applied on each of the difference images where pixels which had intensities above a certain value on either side of zero were masked and chosen for further processing. As a result, two different binary masks were generated: one which comprised of pixels that satisfied \(I_{\mathrm{USM}} > 2.5\sigma\) for RREs and the other one which satisfied \(I_{\mathrm{USM}} < -1.5\sigma\) were considered for RBEs, where \(I_{\mathrm{USM}}\) is the intensity of the difference image post unsharp masking (Fig.~\ref{fig:detect_spicules}~b). The difference in the threshold is due to the skewness in the distribution of RREs and RBEs in the difference maps. In both the masks, pixels which satisfied the thresholding criterion were assigned a value of 1 while the remaining was assigned as 0. Once the binary masks were generated, a morphological opening followed by a closing operation was applied to each of the masks (independently for the RBEs and RREs), on a per time step basis, with a $3\times3$ diamond-shaped structuring element. We refer the reader to \citet{2021A&A...647A.147B} and appendix A.2 of \citet{2021arXiv211010656B} for more details on these morphological operations and the associated reasoning behind them.

Finally, connected component labeling in 3D \citep[i.e. combining both spatial and temporal dimensions, see][]{rosenfeld1966sequential} was performed on the morph processed images so that the RBEs and RREs can be uniquely identified based on a given heuristic. Basically, this technique allows connected neighboring pixels in spatio-temporal domain to be uniquely identified (labeled). To not bias for a particular direction, we employed a 26-neighborhood connectivity criterion in 3D space for this purpose. In other words, two pixels were ``connected" if they shared either an edge, a face or a corner. Furthermore, to avoid erroneous detections and focus primarily on the elongated spicular structures a lower cutoff length of $\sim$200~km (or 8 CHROMIS pixels) was also imposed on the labeled events.  

The above recipe led to a detection of 6457 uniquely labeled events (3623 as RREs/downflowing RREs and 2834 as RBEs) in the complete dataset lasting 11~min over the whole FOV. The occurrence of these (combined) events is shown in the form of 2D probability density map in Fig.~\ref{fig:detect_spicules}~(c) against a background of temporally averaged \hbeta{}~WB image. 

\subsection{Enhancing the AIA images}
\label{subsection:AIA_enhance}

\begin{figure*}[ht!]
% \plotone{Context_CBP_v2.png}
\includegraphics[width=\linewidth]{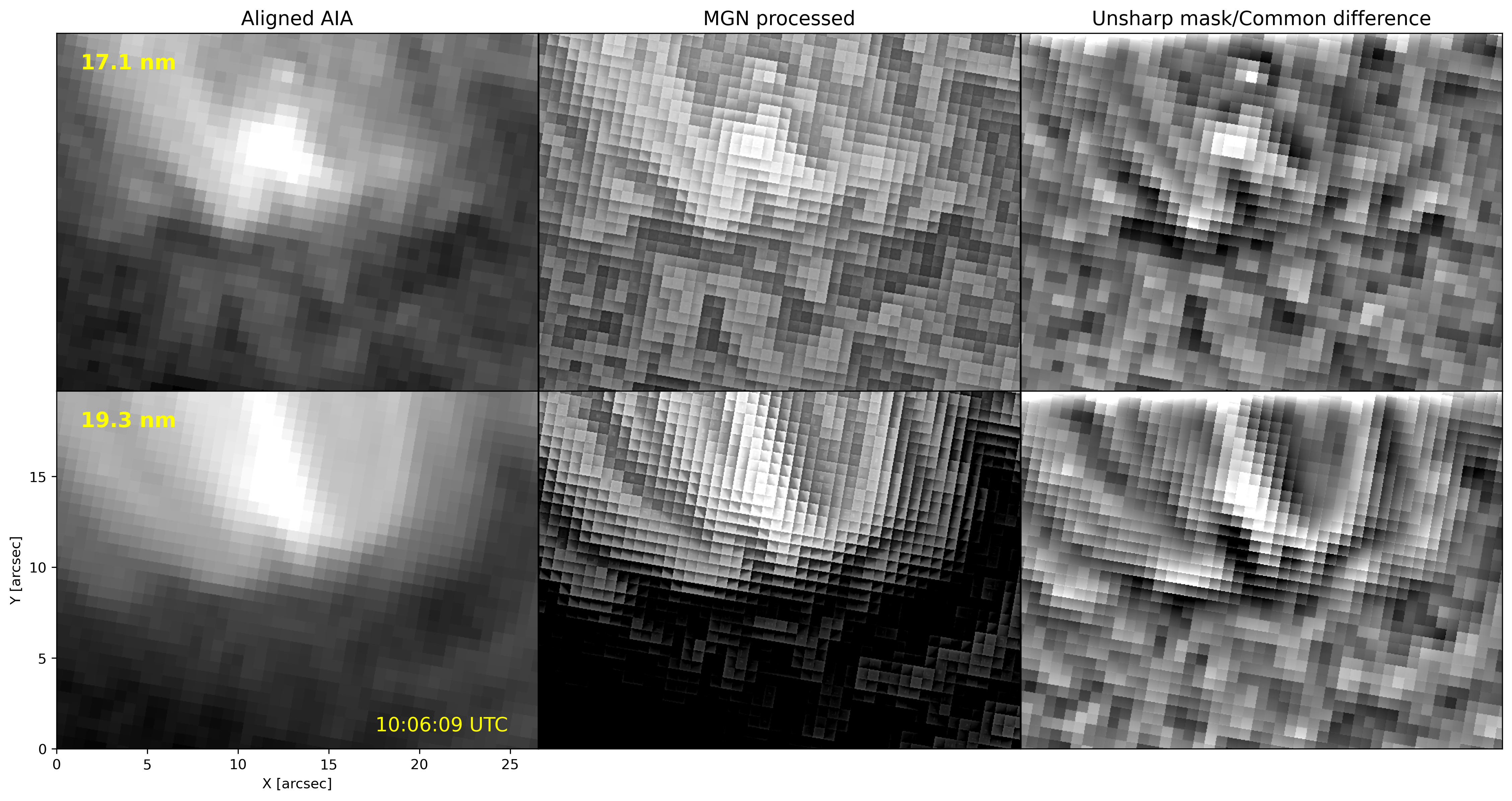}
\caption{Methods of enhancing the AIA images. The top row (from left to right) shows a zoom in to dashed FOV of AIA 17.1~nm intensity map indicated in Fig.~\ref{fig:context} panel~(c), the MGN processed version of the same, and the result of applying the modified common difference technique (see text for details) to the original AIA map, respectively. Bottom row (from left to right) illustrates the result of applying the two enhancement techniques to AIA 19.3~nm channel in the same format as the top row. An animation of this figure is available \href{https://www.dropbox.com/s/0w0noeepegp44ir/Image_comparison.mp4?dl=0}{online}, which shows a comparison between the different enhancement techniques along with the temporal evolution of the disturbances propagating along the loops. The animation shows solar evolution over 11 min.  \label{fig:AIA_enhance}}
\end{figure*}

To facilitate a better understanding of the dynamic relationship between the chromospheric and coronal counterparts of a CBP, it is crucial to enhance the visibility of the coronal images and the loops (strands) associated with the CBP. In this regard, the re-sampled (to CHROMIS pixel scale) AIA images, like the ones shown in the leftmost column of Fig.~\ref{fig:AIA_enhance}, are subjected to a modified version of the common difference technique where the temporal average, over the entire 11~min duration, of each AIA channel is subtracted from an unsharp masked image of the same channel for each time step. This procedure results in images where small changes in the intensity are visibly more enhanced-- due to unsharp masking which adjusts the contrast of the edges (see the rightmost column of Fig.\ref{fig:AIA_enhance}). In addition, the AIA images are also subjected to a multi-scale Gaussian normalization (MGN) procedure \citep{2014SoPh..289.2945M} that enables a better visualization of the overall topology and the orientation of the overlying coronal structure which is not very prominent in original (resampled) AIA images. They are shown in the middle column of Fig.~\ref{fig:AIA_enhance}. The various AIA channels used in this study are MGN enhanced by using the default (same) values of weights and coefficients as in \citet{2014SoPh..289.2945M}.

The animation associated with Fig.~\ref{fig:AIA_enhance} provides a better idea of the advantage of employing the two methods described above and further adds to their comparison with the original co-aligned AIA images. We immediately notice an improvement over the coronal images shown in the left column, where the loops associated with the CBP are barely noticeable. Consequently, the variation in the intensity of the CBP associated with rapid spicular dynamics is shown with the common difference images while the MGN processed images are used as a proxy of the intensity variation in the CBP for all subsequent analysis and results described in this paper. However, it is important to note that MGN does not preserve the photometric accuracy of the images and creates a standardized emission, which is enhanced (subdued) in the regions with lower (higher) intensity. This, however, does not impact the analysis presented in this paper. 

\section{Results and Discussion}
\label{section:results_discussions}
This section presents a detailed description and discussion of the results obtained from the analysis. We begin by investigating the chromospheric foootpoints of the CBP in Sect.~\ref{subsection:cbp_footpoints}, followed by a description of representative examples highlighting the spicule-CBP relationship in  Sects.~\ref{subsection:examples_spicule} and \ref{subsection:twists}. Finally, in Sect.~\ref{subsection:flux_emergence}, we discuss the impact of two small-scale photospheric flux emergence episodes in the chromosphere and the hotter AIA channels.

\subsection{The chromospheric "footpoints" of the CBP}
\label{subsection:cbp_footpoints}

\begin{figure*}[ht!]
% \plotone{Context_CBP_v2.png}
\includegraphics[width=\linewidth]{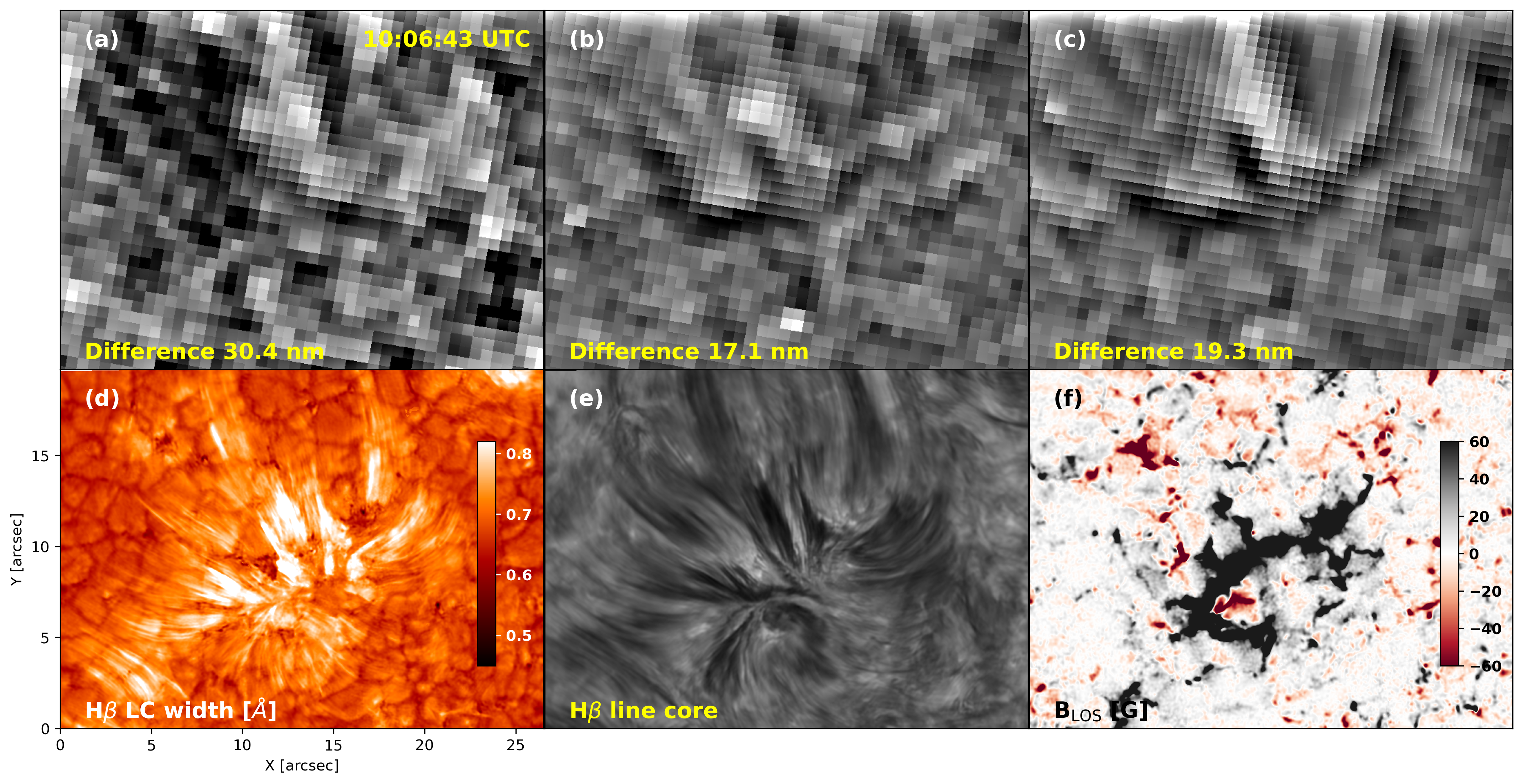}
\caption{The chromosphere and the photosphere underneath the CBP. Panels~(a)--(c) show common difference images in the AIA 30.4, 17.1 and 19.3~nm channels at 10:06:43UT. Panel~(d) shows the co-temporal \hbeta{} LC width map saturated between 0.45--0.82~\AA. Panel~(e) shows the co-temporal \hbeta{} LC image and panel~(f) depicts the corresponding photospheric B$_{\mathrm{LOS}}$ map saturated between $\pm$~60~G. An animation of this figure is available \href{https://www.dropbox.com/s/dcgbc42vtnf6qln/spicule_cbp_overview.mp4?dl=0}{online}, which shows the temporal evolution of the chromospheric and photospheric scenery underneath the CBP for the entire duration of 11 min.  \label{fig:spicule_cbp_overview}}
\end{figure*}

The \hbeta{} wing and the line core images, shown within the dashed FOV in panels~(e)--(g) of Fig.~\ref{fig:context}, depict the chromospheric scene underlying the CBP. The images clearly show multiple dark, elongated, and threaded structures that resemble spicules (or mottles). A zoom-in to the dashed FOV is shown in Fig.~\ref{fig:spicule_cbp_overview} which focuses solely on the region in and around the CBP. To aid better visualization of the intensity disturbances propagating in the CBP, we show the common difference images for the different AIA channels in panels~(a) through (c). Panel~(d) shows the \hbeta{} line core (LC) width map which is basically the wavelength separation at half the intensity range between the minimum of the \hbeta{} line profile and the average intensities at a displaced wing position from the line center \citep[following][]{2009A&A...503..577C} for each pixel on the FOV. In this case the displacement parameter was set at $\pm$~66~pm from the line center which was determined by converting the displacement parameter of 90~pm for the \halpha{} spectral line, chosen by \citet{2009A&A...503..577C}, into equivalent Doppler units (\kms{}). RBEs and RREs (including downflowing RREs) appear to be in ``emission" (compared to the background features as seen in panel~d) in these maps since they generally have enhanced opacity owing to their broad LOS velocity distribution \citep[][]{2016ApJ...824...65P,2021A&A...647A.147B} and enhanced temperature \citep{2012ApJ...749..136L}. Panels~(e) and (f) show the co-temporal \hbeta{} line core intensity and the LOS photospheric magnetic field maps underneath the CBP. Spicules and/or mottles dominate the whole FOV and they are seen to be predominantly rooted in the close vicinity of the strong (negative) polarity magnetic field patch which also happens to be the photospheric magnetic roots of the CBP.

%A similar picture was also outlined in \citet{2021A&A...646A.107M}, however, due to the significantly higher resolution of our dataset the appearance and changes within a few seconds were never reported.  The animation associated with the figure 

 A glance at panels~(b)--(e) of Fig.~\ref{fig:spicule_cbp_overview} immediately suggests that the CBP loops and their chromopsheric counterparts bear a close morphological resemblance. This is further highlighted in the animation associated with the figure where the 17.1 and 19.3~nm loops appear to have propagating disturbances nearly in tandem with the rapid changes in the chromosphere, especially toward the later half of the data sequence. The 30.4~nm common difference image appears to be noisier and it does not show the loops associated with the CBP as prominently as in the other AIA channels. However, the animation shows clear disturbances associated in the same region as underlying spicules but the overall morphology is less pronounced (compared to panels~b and c) making them difficult to relate visually. The lack of loop-like appearances in the 30.4~nm channel could be attributed to its relatively lower temperature sensitivity (log~$T$(K)~$\sim$~4.7) compared to 19.3 and 17.1~nm channels which have a peak temperature sensitivity of around log~$T$(K)~$\sim$~6 \citep[][]{2012SoPh..275...41B}. Moreover, it is rather common to observe the relatively cooler footpoints of the CBPs in the 30.4~nm channel underneath the hotter loops \citep{2012ApJ...757..167K,2019LRSP...16....2M,2021A&A...646A.107M} which may further justify the less pronounced morphological resemblance. 
 
\citet{2021A&A...646A.107M} report that the chromospheric counterpart of a CBP largely comprises of elongated, dark features when observed in the \halpha{} line core images. They name these features ``\halpha{} loops" which also appear to constitute a fundamental part of the overall magnetic structure of the CBPs. While we do not find the existence of such loops likely due to our observations being limited to a part of the entire CBP (thereby missing the opposite polarity), spicules dominate our FOV and plays a central role in driving the dynamics of the chromosphere underneath the CBP. 
%  Our analysis is in line with their findings since it is possible that many of the spicular features that we observe in our dataset would be described as mottles under lower spatial and temporal resolution \citep[see][for the effect of spatio-temporal resolution on deduced spicule properties]{2013ApJ...764...69P}. 
%  Moreover, the current work focuses primarily on investigating a spicular connection in the chromosphere underneath the CBP with its coronal counterpart at high spatial and temporal scales that have never been reported before. 
 
 \begin{figure*}[ht!]
% \plotone{Context_CBP_v2.png}
\includegraphics[width=\linewidth]{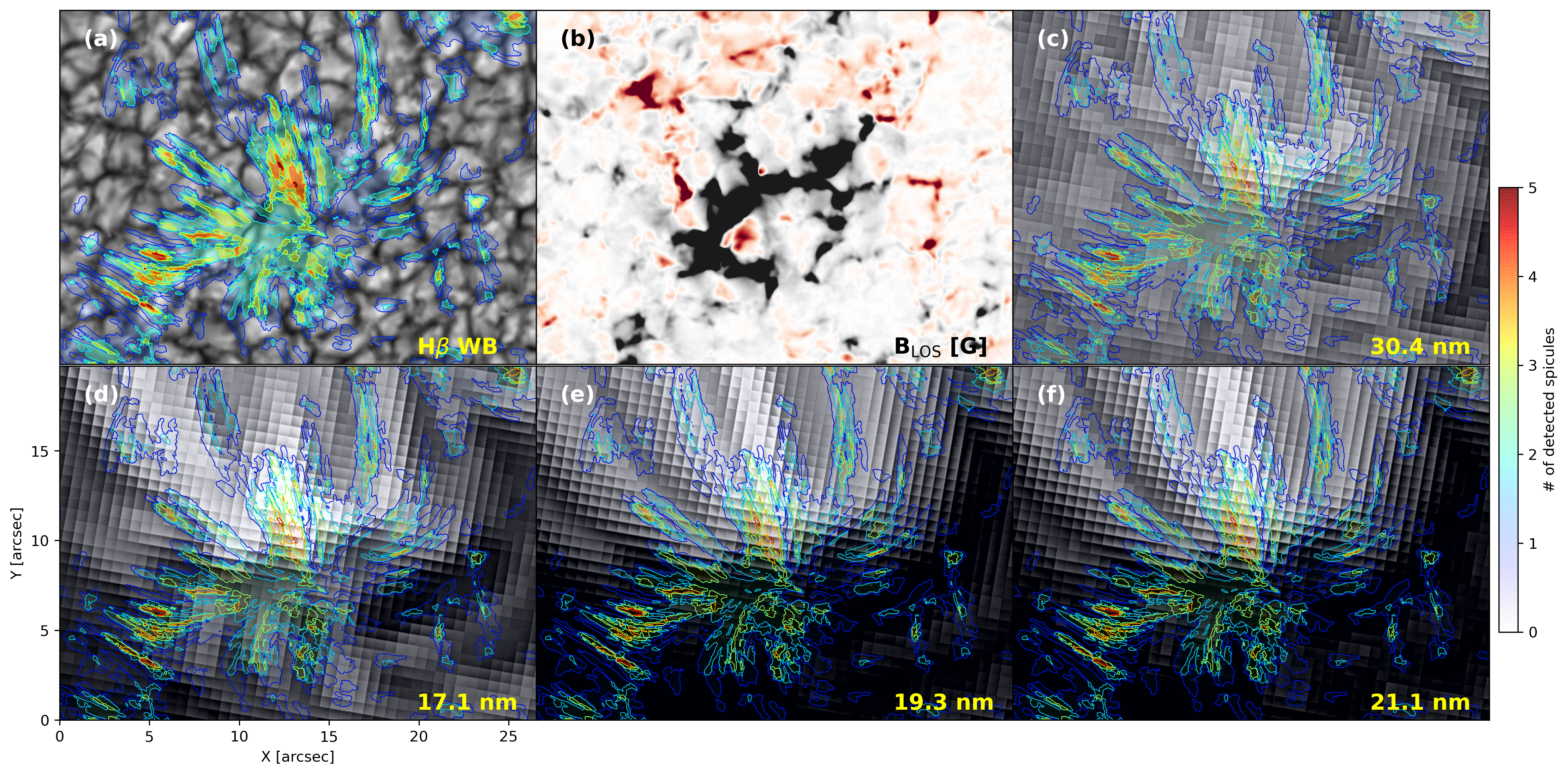}
\caption{Morphological similarities between spicules and the loops associated with the CBP. Panels~(a) and (c)--(f) show the 2D density map of the detected spicules overlaid against a background of temporally averaged \hbeta{} WB, MGN enhanced AIA 30.4, 17.1, 19.3, and 21.1~nm channels, respectively, whereas panel~(b) shows a temporal average of the underlying photospheric magnetic field saturated between $\pm$~60~G. The FOV in each of the panels correspond to the dashed FOV indicated in Fig.~\ref{fig:context}. \label{fig:spicule_2D_distribution}}
\end{figure*}

Figure~\ref{fig:spicule_2D_distribution} shows the occurrence of the detected on-disk spicules, using the method described in Sect.~\ref{section:methods}, in the form of a 2D density map against a background of temporally averaged images for four MGN processed AIA channels (panels~c--f) and an SST WB channel (\hbeta{}, panel~a). As described in Sect.~\ref{subsection:AIA_enhance}, the MGN processed images show the intensities in absolute units (though it fails to preserve the photometric accuracy) unlike the common difference images. From this figure it is clear that the distribution of spicules is very well correlated with the orientation and overall morphology of the CBP loops, as is evident from the 17.1, 19.3 and 21.1~nm channels. This provides a compelling observational confirmation (in a statistical sense) of spicules tracing the coronal magnetic field lines which, to the best of our knowledge, has not been reported before. Moreover, we also notice that the number density of the detected spicules is predominantly located close to the footpoint of the CBP loops. This scenario seems to suggest that the studied spicules are the cromospheric components of the CBP loops which, post heating, appear in the hotter TR and coronal channels, and further contribute toward the transient intensity disturbances in the already hot CBP loops.\citep[see for example][and the references therein for studies about the coronal counterpart of spicules]{2011Sci...331...55D,2011A&A...532L...1M,2014ApJ...792L..15P,2015ApJ...799L...3R,2017ApJ...845L..18D,2019Sci...366..890S}. We will explore this aspect further with a few representative examples in Sect.~\ref{subsection:examples_spicule}.

The morphological similarities between the \hbeta{} spicules and the coronal loops associated with CBP indicates the possibility that the loop structures are associated with spicular mass ejections and transient heating of the plasma from chromospheric to coronal temperatures. A direct investigation of such a connection would however require a more detailed analysis by combing high-resolution numerical simulations with spectroscopic observations of the CBP. Nonetheless, some studies such as \citet{2017ApJ...845L..18D} already showed an intriguing connection between spicules in the TR and the formation of coronal strands in a decayed plage region with the help of numerical simulations and coordinated IRIS and SDO observations. Moreover, spicules were also found to be responsible in triggering propagating coronal disturbances (PCDs) along many of the pre-existing (and newly formed) coronal strands rooted to the plage. PCDs are rapid recurring intensity fluctuations ($\sim$~100~\kms{}) whose exact nature remains a mystery, especially outside of the sunspots \citep[see][for example, on the discussion whether PCDs are flows or waves]{2010ApJ...722.1013D,2009SSRv..149...65D,2015SoPh..290..399D,2016ApJ...829L..18B}. Therefore, it is likely that the intensity disturbances observed in the common difference coronal images are linked to the rapid spicular dynamics in the chromosphere.

From Fig.~\ref{fig:spicule_2D_distribution} we also notice a significant overlap between the widths of the detected chromospheric spicular features and the observed loops associated with the CBP. Using coordinated observations from Hinode's Extreme ultraviolet Imaging Spectrometer and Transition Region and Coronal Explorer instruments, \citet{2009A&A...497..287D} derived the volumetric plasma filling factor in CBPs, and came to the conclusion that the widths of its loops can be between 0\farcs2--1\farcs2 with possible substructures that are below the resolution limit of the instruments. Comprehensive statistical analysis carried out by \cite{2012ApJ...759...18P} and \citet{2021A&A...647A.147B}, indicate that spicule widths, for both off-limb and on-disk cases, are consistent with the range reported by \citet{2009A&A...497..287D} which further suggests that the \hbeta{} spicules detected in this study are likely the chromospheric counterparts of the CBP. 

Numerical modeling efforts led by \cite{2018ApJ...860..116M} offer key insights into the role of spicules in determining the widths of the coronal loops. They report that the widths of the simulated spicules (and subsequently the coronal loops) are primarily determined by the driving mechanism that generates these flows, along with the overall magnetic topology and heating within the magnetic field lines. Moreover, they find that the magnetic field rapidly expands primarily between the photosphere and middle to upper chromosphere where spicules are seen to be generated (in the model). The expansion of the field line is rather insignificant between the transition region and the corona which may explain why the CBP loops and spicules appear to have similar widths. 

% Furthermore, Joule heating and thermal conduction were found to be responsible for providing mass and energy flux to the corona which could also be the case with the CBP loops. Despite the above explanations, it is important to remark that the numerical simulations carried out by \citet{2018ApJ...860..116M} are restricted to 2D. It remains to be seen how these interpretations are impacted or modified for a 3D simulation.

\subsection{Representative examples of spicular-CBP connection}
\label{subsection:examples_spicule}

\begin{figure*}
\plottwo{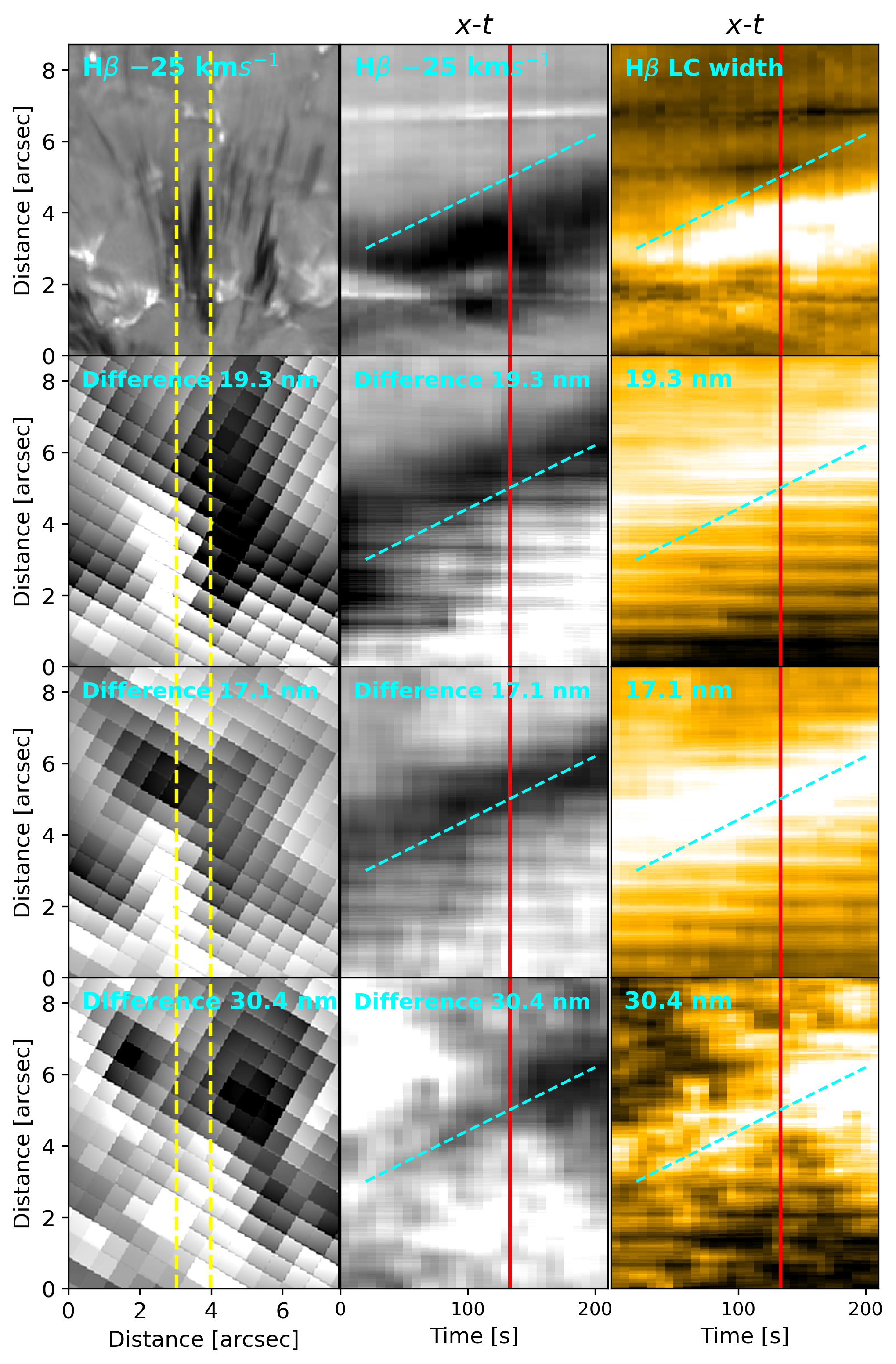}{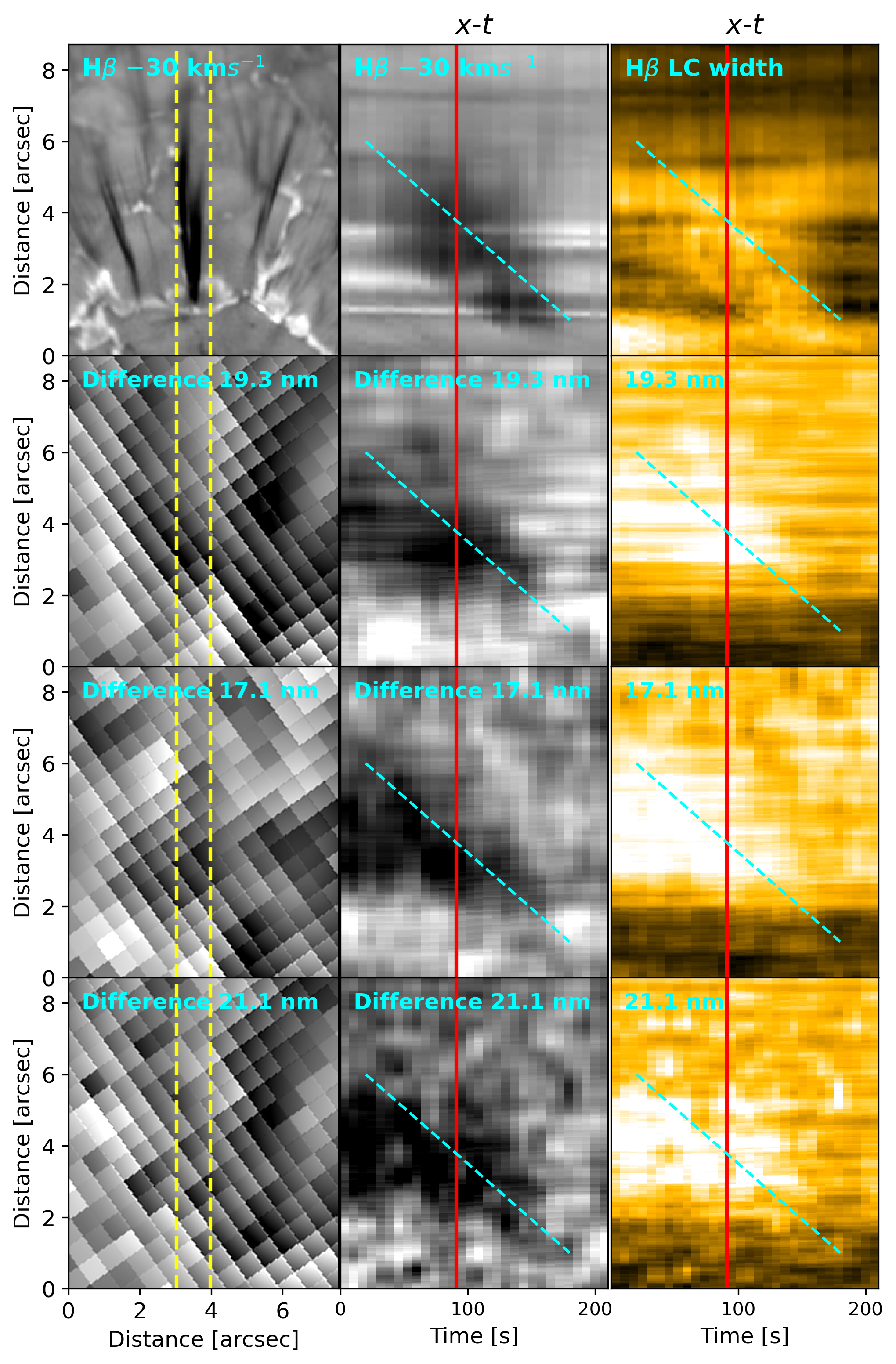}
\caption{Two representative examples highlighting the spicule-CBP connection from the chromosphere to the corona. Left panel: an example of an RBE observed in the blue wing ($-$~25~\kms{}) of \hbeta{} spectral line and its associated propagation in the different AIA passbands as indicated. The dashed vertical lines in the leftmost column
indicate the region along which the \xt{} maps have been extracted for the different channels as shown in the middle and the rightmost columns. The solid vertical red lines in the \xt{} maps correspond to the instant at which this figure is shown. The dashed cyan line serves as a reference to illustrate the direction of propagation. Right panel: another example of a spicule observed in the blue wing ($-$~30~\kms{}) of the \hbeta{} spectral line is shown along with its impact on the AIA channels in the same format as the left panel. Note that the apparent direction of propagation of this spicule is opposite to the example presented in the left panel. Animations of the two panels are available \href{https://www.dropbox.com/sh/i7xdzzhbcmos6cd/AAAC1riVB9Wmgh7pr_csOZDsa?dl=0}{online}.
\label{fig:spicule_example} They show the spatio-temporal evolution of the two spicules in the chromospheric \hbeta{}, transition region 30.4~nm and coronal 17.1, 19.3 and 21.1~nm channels during their respective lifetimes.} 
\end{figure*}

In this section we further illustrate the spicule-CBP connection discussed in Sect.~\ref{subsection:cbp_footpoints} through two representative examples shown in the left and right panels of Fig.~\ref{fig:spicule_example} including their signatures in the TR (AIA 30.4~nm) and coronal passbands. We show the common difference images for the different AIA channels (in the left columns of each of the two panels) in order to enhance the visibility of the changes in intensity. The dashed vertical yellow lines in the left columns of both panels show the region of interest that is chosen to construct the \xt{} maps. Moreover, in addition to the common difference, we also show the \xt{} maps derived from the MGN processed AIA images and \hbeta{} LC width maps to highlight the temporal evolution of the plasma emission from each channel.  

The left panel shows an example of an RBE in the blue wing of \hbeta{} (at $-25$~\kms{}). From the animation and the \xt{} maps (top row), it is clear that the RBE has an outward (away from the bright network regions) apparent motion and propagates from $\sim$~2\arcsec\ to 6\arcsec\ in the vertical direction during its evolution. This is a commonly observed property of spicules where they originate from strong magnetic flux concentrations and tend to shoot outwards. Since spicules often have a wide range of Doppler shifts associated with them \citep{2016ApJ...824...65P,2021A&A...647A.147B}, analysis based on images at fixed wavelength positions can sometimes provide an incomplete picture of their evolution. In such cases LC width maps offer a better understanding since they are determined by considering a range of wavelengths on either side of the line center (see Sect.~\ref{subsection:cbp_footpoints}). In the present example however, the \xt{} maps derived from the LC widths and \hbeta{} blue wing images are seen to be well correlated with each other. %%%%%-------------- the following sentence should go in the introduction ------------
%%%%%It is important to note here that on-disk spicules have traditionally been imaged in the \halpha{} \citep{}
%%%%%-------------- this sentence should go in the introduction ------------

A comparison of the spatio-temporal evolution seen in the corresponding AIA channels show a noticeable correlation with the \hbeta{} counterpart. An inspection of the animation of the AIA difference images shows clear intensity disturbances propagating in the CBP which appear to be in tandem with the \hbeta{} spicule. The 19.3 and 17.1~nm difference images, in particular, show a clear propagation from the bottom to the top of the FOV. This is also well highlighted in the difference \xt{} maps (middle column). The 30.4~nm difference images, on the other hand, does not show such a clear propagating disturbance however, the \xt{} maps reveal clear signatures which are also in tandem with the other wavelength channels.

A close look at the different \xt{} maps associated with the left panel of Fig.~\ref{fig:spicule_example} reveals a small but distinct spatial (and/or) temporal offsets among the different channels (with respect to the dashed cyan line) -- with the TR and coronal emission lying above the cooler chromospheric plasma. Such a scenario is consistent with the analysis presented by \citet{2011Sci...331...55D, 2014ApJ...792L..15P}, and it suggests that the RBE has a multi-thermal nature with temperatures that can range from chromospheric to coronal temperatures (of at least 1~MK). In fact, an early study focusing on multi-wavelength diagnostics of a CBP by \citet{1981SoPh...69...77H}, found that coronal emission in CBPs lie a few arcseconds over an above the chromospheric emission suggesting the hypothesis that magnetic loops in a CBP are rooted in the chromosphere. The spatial offset between the TR (30.4~nm) and coronal (17.1/19.3~nm) emission patterns is indicative of the fact that the emission in the coronal channels are not caused by relatively cooler ions \citep[such as \ion{O}{v}, see][]{2012SoPh..275...41B} which are sensitive to temperatures of about 0.2~MK under equilibrium conditions. Moreover, the emission from the \ion{O}{v} is expected to be very faint in comparison to the dominant \ion{Fe}{ix} and \ion{Fe}{xv} ions and would have occurred in the same spatial region as the 30.4~nm emission. This further adds support in favor of the spicular contribution to coronal emission associated with the CBP.

The right panel of Fig.~\ref{fig:spicule_example} shows another example of spicular connection associated with the CBP in the same format as the previous example. Unlike the left panel, the spicule appears to be downward propagating as is evident from the animation and the \xt{} maps in the middle and right columns. A quick glance at the \hbeta{} image would suggest that the example here is a blue-wing counterpart of downflowing RREs \citep[seen in the red wing images of \halpha{}, see][]{2021A&A...647A.147B,2021A&A...654A..51B}. However, a closer inspection of the animation reveals a rather complex scenario where the spicule is rapidly seen to change its orientation (with respect to the LOS of the observer) during its propagation, before finally disappearing around $t=180$~s. Such a complex propagation seems to convey that the spicule is downward propagating, which in reality could be the opposite.

Regardless of the interpretation associated with the orientation of the spicule and its mass flow, interestingly (and more importantly), the \xt{} maps of the coronal channels show a remarkable correlation with \hbeta{}. Moreover, the spatial offsets among the different channels are consistent with the discussion presented in the previous example and conforms with the multi-thermal aspect of the spicule and its relation to the CBP. This supports our proposition that spicules in the chromosphere have a direct relationship with the disturbances propagating in the CBPs. 
Although many questions remain, this may also provide support to the idea that the processes associated with spicules may play a role in providing  mass and energy flux necessary to sustain the radiative and conductive energy loses in the solar corona as suggested in the numerical simulation studies by \citet{2017Sci...356.1269M,2018ApJ...860..116M}.

% \textbf{From the current set of observations, it is not straightforward to investigate whether spicules observed at the footpoints of the CBP are unique in nature. However, it is expected that the strength of the magnetic field (and its inclination) in the lower atmosphere plays a major role in driving the observational properties of spicules. Statistical analysis by \citet{2012ApJ...759...18P} reveal clear differences between properties of spicules in active regions and quiet-Sun (and coronal holes), and \citet{2011ApJ...743..142H} report similar findings with numerical simulations. Underneath the CBP, the field strength is distinctly stronger compared the rest of the FOV--where the rapid expansion of the weaker field lines likely leads to different coronal impact. Statistical studies with coordinated chromospheric and (higher-resolution) coronal observations are needed to determine if the coronal contribution of spicules depend on the strength of the photospheric magnetic fields.  }

\subsection{Twists at the footpoints of the CBP}
\label{subsection:twists}

\begin{figure} 
    % \epsscale{0.85}
    \plotone{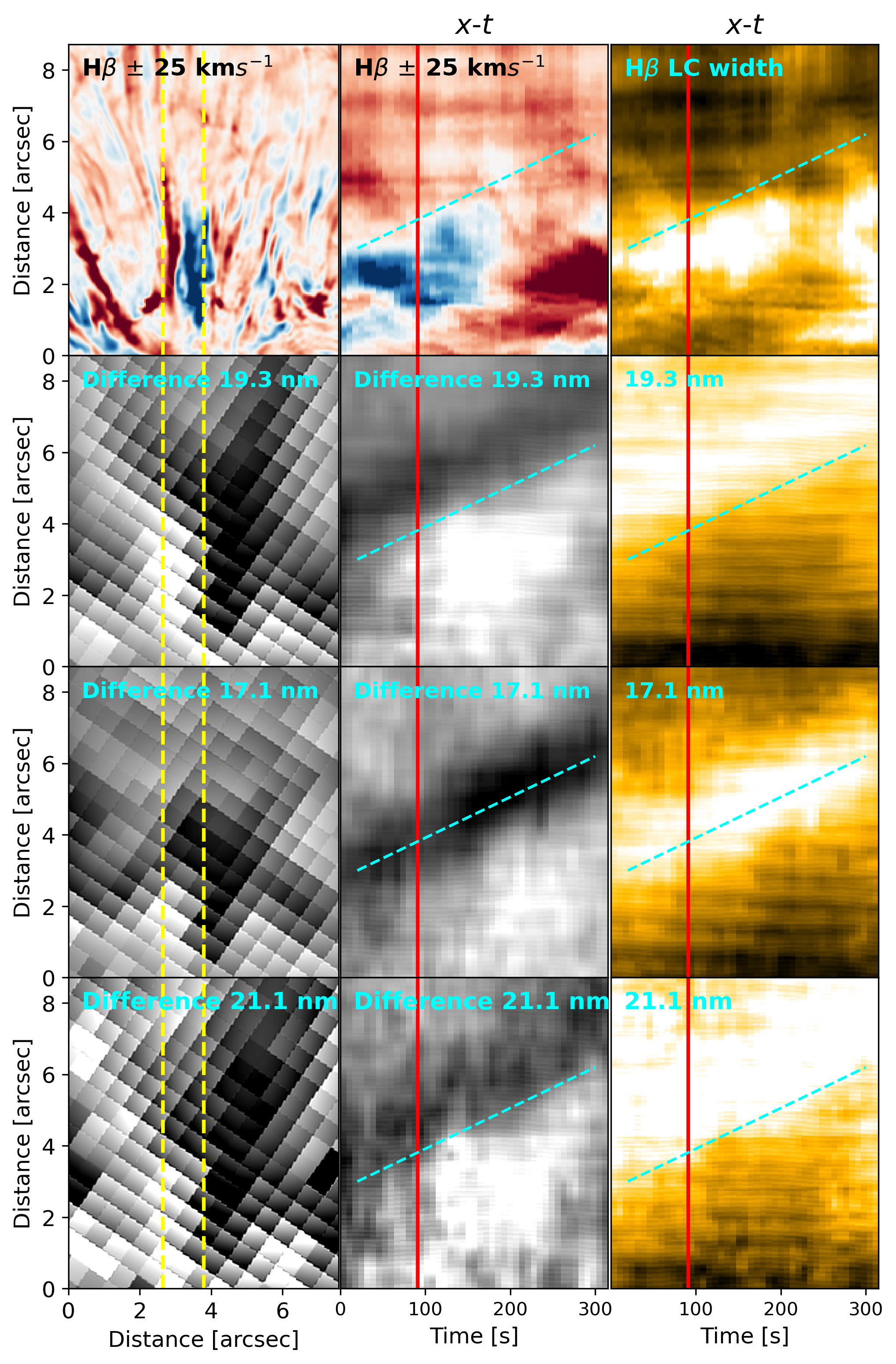}
    \caption{Chromospheric twist and its likely propagation into the solar corona. This figure is in the same format as Fig.~\ref{fig:spicule_example} except that the \hbeta{} wing images is replaced by its Dopplergram at $\pm$~25~\kms{}. Blue (red) color is indicative of plasma motion toward (away from) the observer. The associated animation (available \href{https://www.dropbox.com/s/utb4pjjz4iwz7rv/CBP_twisting_different_color.mp4?dl=0}{online}) shows the spatio-temporal evolution of the twist propagation across the chromospheric and coronal channels for roughly 300~s.  \label{fig:twisting_example}}
\end{figure}

Spicules are known to undergo twisting (torsional) motions that are often interpreted as a sign of Alfv\'enic waves responsible for driving the fast solar wind and balancing the energy losses suffered in the solar corona \citep{2007Sci...318.1574D,2011Natur.475..477M,2012ApJ...752L..12D}. Moreover, small-scaled twists associated with spicules are ubiquitously found in the solar atmosphere (active regions and quiet Sun alike), and their signatures have also been found in the TR \citep{2014Sci...346D.315D}. 

In Fig.~\ref{fig:twisting_example}, we show a case of twist associated with spicules present at the footpoints of the CBP and their influence on the coronal loop above. The top row of the figure shows an \hbeta{} Dopplergram at $\pm$~25~\kms{} with blue and red colors indicating plasma motions toward and away from the observer along the LOS. The corresponding \xt{} map and the animation shows a clear change of direction (or color from blue to red) indicating a definite twisting motion in the chromosphere. A close look at the animation indicates that the event starts with a predominantly positive (red) Doppler shift at $t=0$~s which rapidly converts to a negative (blue) Doppler shift in roughly 30~s. It remains predominantly negative until around $t=180$~s after which it rapidly twists toward positive (red) once again. This behavior is very similar to the examples observed in the \halpha{} and \ion{Ca}{ii}~854.2 spectral lines for both off-limb and on-disk spicules as outlined in \citet{2012ApJ...752L..12D} and \citet{2014Sci...346D.315D}. The propagation speed (of roughly 35~\kms{}) is fairly consistent (given the uncertainty related to  the viewing angle) with Alfv\'en speeds at chromospheric heights \citep{2007Sci...318.1574D}. The LC width \xt{} map shows a similar trend as the Dopplergram \xt{} map where additionally we see that the plasma associated with the twisting motion stands out distinctly with respect to the background. 

The \xt{} maps associated with the difference and MGN enhanced AIA 19.3, 17.1 and 21.1~nm channels show strong emission which are in tandem with their chromospheric counterpart (similar to the examples shown in Fig.~\ref{fig:spicule_example}). We also notice a significant offset in the emission among the different channels indicating that the plasma is heated to temperatures of at least 1--2~MK (refer to the discussion in the previous section), in association with the twisting spicules at chromospheric temperatures. Of course, a complete analysis of such a twist propagating in the coronal loops necessitates spectroscopic studies of the solar corona which is not possible with the set of instruments used in this study. Moreover, current observations suggest the prevalence of Alfv\'enic waves in the corona \citep[e.g.,][]{2007Sci...317.1192T,2011Natur.475..477M} although the wave energy flux and wave modes are poorly captured with current instrumentation. Alfv\'enic waves have the potential to heat coronal loops \citep{2018ApJ...856...44A}, in particular when mass flows are present within a structure in addition to waves \citep[as shown for example in][]{2009ApJ...694...69T,2016ApJ...817...92W}. Upcoming space missions, such as Solar-C/EUVST or the MUlti-slit Solar Explorer \citep[MUSE,][]{2020ApJ...888....3D,2022ApJ...926...52D}, can potentially address these aspects. 

\subsection{Chromospheric and coronal response to emerging magnetic flux}
\label{subsection:flux_emergence}

\begin{figure*}[ht!]
% \plotone{Context_CBP_v2.png}
\includegraphics[width=\linewidth]{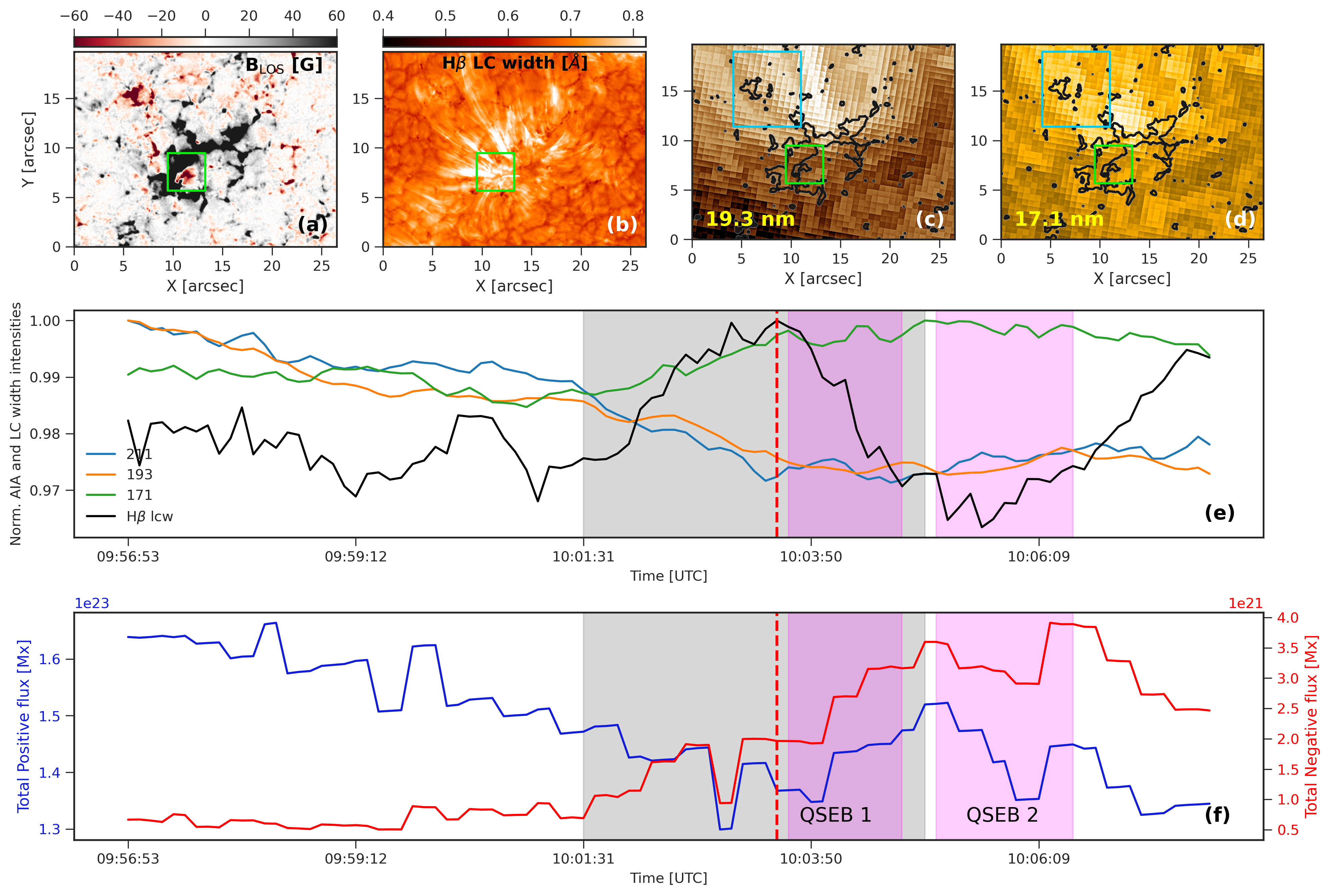}
\caption{Chromospheric and coronal response to flux emergence episode 1. Panel~(a) shows the photospheric LOS magnetic field map, (b) shows the \hbeta{} LC width map, (c) and (d) shows the overlying corona of the CBP in AIA 19.3~nm and 17.1~nm channels. The green square boxes drawn in panels~(a)--(d) show the region associated with the flux emergence event. Panels~(c) and (d) also shows a blue rectangular box centered around ($X$,$Y$)=(7\farcs5,15\arcsec) where the coronal response to the emerging flux is analyzed. The black contour in panels~(c) and (d) indicates the regions with
an absolute LOS magnetic field $\geq$~50 G. Panel~(e) shows the chromospheric and coronal light curves obtained from the green and blue FOVs indicated in panels~(b)--(d), respectively. Panel~(f) shows the temporal variation of the total positive and negative magnetic flux in the region bounded by the green colored box in panel~(a). The gray shaded intervals in panels~(e) and (f) shows the time when the chromospheric spicular activity is enhanced, and the pink shaded regions indicates the interval when the two QSEBs are observed. Animation of this figure is available \href{https://www.dropbox.com/s/r5tk66yapaytjcw/FE_CBP_spicule_FOV3_flux.mp4?dl=0}{online}, which shows the evolution of the magnetic flux and its response in \hbeta{}, AIA 19.3 and 17.1~nm channels in the form of light curves for the entire 11~min of solar evolution. \label{fig:flux_emer_1}}
\end{figure*}

In this section, we investigate the chromospheric and coronal responses to two emerging flux episodes as observed from the LOS magnetic field.

\subsubsection{Flux emergence episode 1}
\label{subsubsection:FE_1}

Figure~\ref{fig:flux_emer_1} and associated animation show an overview of the first episode. In the LOS magnetogram of panel (a), a negative parasitic polarity is seen to emerge in a predominantly positive region. In the figure, the area where this episode takes place is bounded by a green square region of 100$\times$100 pixels. The variation of the total positive and negative magnetic fluxes within this green square is shown in panel~(f) where we find that the total negative flux starts to increase steadily after 10:01:31UT, reaching its peak value of $\approx$4$\times$10$^{21}$~Mx around 10:06:30 UT. 

We investigated the subsequent chromospheric response to the flux emergence episode by analyzing the \hbeta{} LC width maps (panel~(b)). Compared to looking simply at the \hbeta{} line core, the LC width is optically thin toward the dense fibrilar canopies visible in the line core images, thereby facilitating a better understanding of the ``connection" between the spicules and their photospheric footpoints. In particular, we obtain the light curve within the green square since changes in the chromosphere and spicules associated with this flux event would likely be rooted near this region. The result is shown as a black curve in panel~(e). There is a clear enhancement in the \hbeta{} LC width intensity starting from $\approx$ 10:01~UT (around the same time as the total negative flux in panel~(f) starts to show a marked increase). The LC intensity continues to increase until $\approx$10:03:30~UT after which it starts to decay and reaches a minimum around 10:05:15~UT, incidentally after which the increase in the total negative flux (by $\approx$300\% from the start of the event) also tends to stabilize.  It seems that as the flux emerges and subsequently interacts with the dominant positive polarity through magnetic reconnection (see Section \ref{subsubsection:reconnection}), there is a significant enhancement in the spicular activity compared to the any of the previous time steps implying a correlation between the two. This is consistent with the analysis presented in \citet{2021A&A...646A.107M}, in the context of chromospheric response to flux emergence associated with a CBP, and also \citet{2019Sci...366..890S}, in general. However, unlike \citet{2019Sci...366..890S}, it is not implied that the flux emergence (and subsequent cancellation) leads to the ``generation" of spicules seen in close proximity. Instead, it is more appropriate to say that the emergence likely caused an enhancement in the observed spicular activity. We notice the presence of spicules both well before and after the emergence event as is evident from the animation. 

We have also investigated the coronal response using the SDO/AIA 19.3 and 17.1 channels (panels (c) and (d)) along with the 21.1~nm channel. In this case, the corresponding light curves, shown in panel~(e), are obtained in a region that is spatially displaced from the emerging flux region (the blue rectangle of 200$\times$300 pixels in panels~(c) and (d)). This is because it was shown earlier on in the paper that spicules lie close to the footpoints of the CBP structure, whereas the coronal loops extend well beyond and are spatially (and/or temporally) displaced. Before 10:01:31~UT, the 17.1, 19.3, 21.1~nm light curves have very similar intensity levels relative to the maximum of each channel. As soon as the chromospheric activity starts to increase from 10:01:31~UT, we notice a co-temporal increase in the intensity of the 17.1, whereas the 19.3 and 21.1 channels show a steady decrease compared to their respective pre-event values. However, unlike the  illustrative spicule examples shown in the previous section (i.e. in  Figs~\ref{fig:spicule_example} and \ref{fig:twisting_example}) and also the many studies conducted in the past \citep[such as,][]{2011Sci...331...55D,2011Natur.475..477M,2016ApJ...820..124H} that do establish a coronal connection quite convincingly in different SDO/AIA channels, for this particular enhanced spicular episode, the coronal relation is not clear. To not bias our interpretation, we have also computed other light curves over different AIA FOVs (of the same area) spatially displaced from one another, finding similar results.

\subsubsection{Reconnection associated with flux emergence 1}
\label{subsubsection:reconnection}
\begin{figure*}[ht!]
% \plotone{Context_CBP_v2.png}
\includegraphics[width=\linewidth]{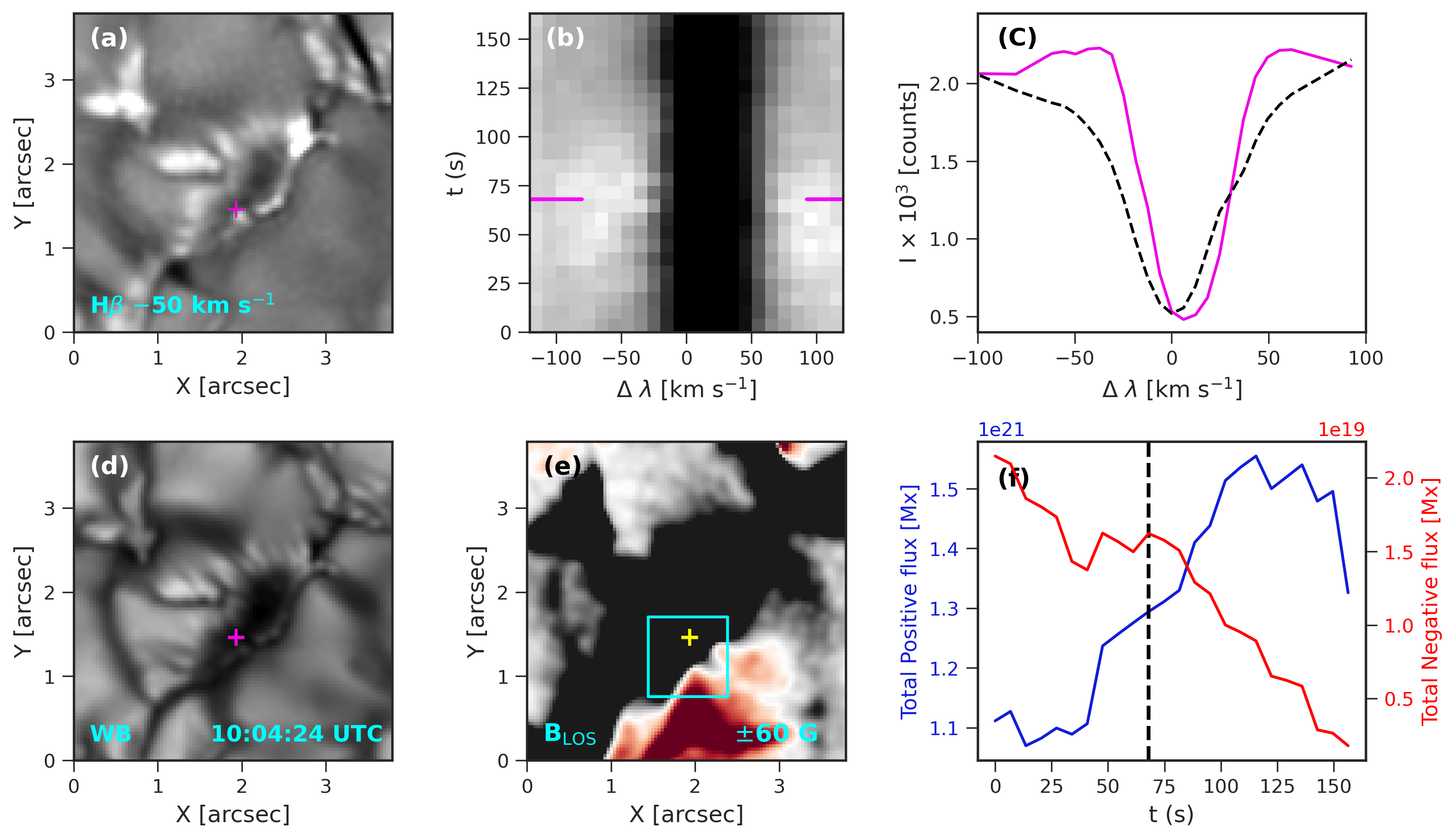}
\caption{Details of QSEB~1 observed during the flux emergence episode 1. Panel~(a): QSEB observed in the far blue wing of \hbeta{}. Panel~(b): temporal variation of the \hbeta{} line profile for a location in the QSEB indicated by the magenta marker in panel~(a) in the form of a $\lambda-t$ diagram. Panel~(c): \hbeta{} spectral line at a temporal instant indicated by the marker in panel~(b) and the spatio-temporal average \hbeta{} reference profile (dashed black line). Panel~(d): corresponding WB image. Panel~(e): corresponding LOS magnetic field map saturated between $\pm$~60~G, and panel~(f) temporal evolution of the total positive and negative magnetic flux within the cyan box shown in panel~(e). The dashed vertical line in (e) indicates the instant when this figure is shown. Animation of this figure is available \href{https://www.dropbox.com/sh/lzfjgxjbnb8rxg4/AADD65UBS7OJPE2OGQ9qvmpXa?dl=0}{online} and it shows the evolution of the magnetic field, the QSEB and the corresponding \hbeta{} spectra before, during and after the appearance of the QSEB for about 2.5~min of solar evolution. \label{fig:QSEB1}}
\end{figure*}

\begin{figure*}[ht!]
% \plotone{Context_CBP_v2.png}
\includegraphics[width=\linewidth]{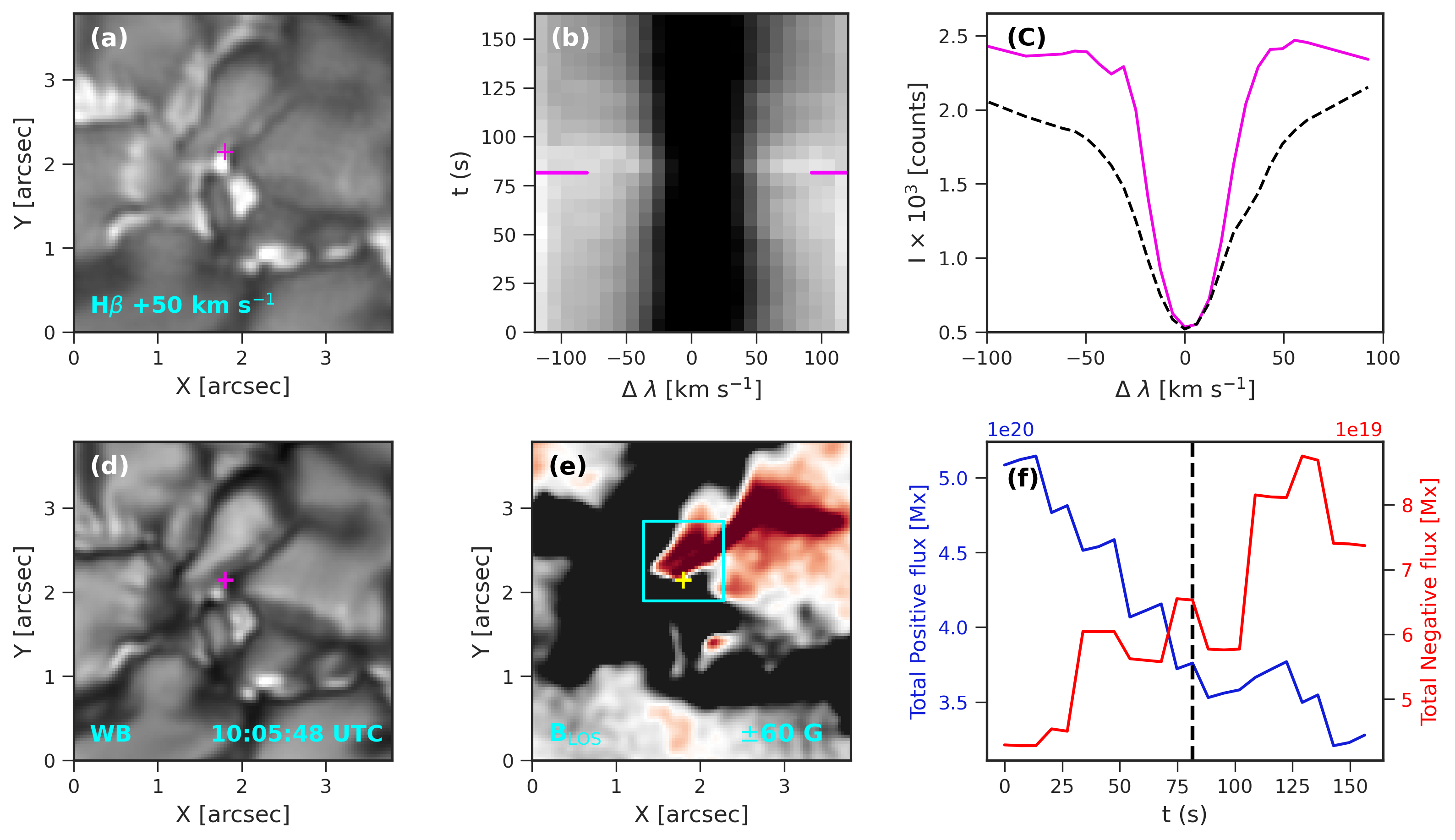}
\caption{Details of QSEB~2 associated with the flux emergence event 1. The figure and its associated animation (available \href{https://www.dropbox.com/sh/lzfjgxjbnb8rxg4/AADD65UBS7OJPE2OGQ9qvmpXa?dl=0}{online}) showing about 2.5~min of solar evolution displays the temporal evolution of the magnetic field, QSEB and the \hbeta{} spectra in the same format as Fig.~\ref{fig:QSEB1}. \label{fig:QSEB2}}
\end{figure*}

%An increase in the coronal emission associated with flux emergence in a CBP is often linked to the phenomenon of reconnection associated heating of the solar corona \citep[see][and references therein]{1981SoPh...69...77H,2016ApJ...818....9M,2018ApJ...864..165W,2021A&A...646A.107M}. In such cases, reconnection in the corona can be thought to be caused by random small-scale footpoint motions which can then efficiently transmit heat along the magnetic field lines via thermal conduction (at temperatures $\geq$~10$^{4.7}$~K). But, is it necessary for the magnetic field to reconnect only in the solar corona? Is it possible that the emerging flux interacts with the ambient magnetic field deeper in the solar atmosphere and cause (smaller-scaled) reconnection close to the footpoints? We investigate such a possibility in this section in the context of Ellerman bombs \citep[EBs,][]{1917ApJ....46..298E}.

Figures~\ref{fig:QSEB1} and \ref{fig:QSEB2} show evidences of magnetic reconnection through two examples of EBs located in the footpoint of the CBP associated with the emerging flux episode described above. We refer to them as quiet-Sun EBs (QSEBs) after \citet{2016A&A...592A.100R} where these were first described. QSEBs are smaller, shorter-lived and less intense brightenings and are found in relatively quieter areas on the Sun compared to their active region counterparts. With the help of high-quality \hbeta{} observations from the SST, \citet{2020A&A...641L...5J,2022A&A...664A..72J} recently showed that QSEBs are ubiquitous in the solar atmosphere and can play an important role in the energy balance of the chromosphere. Recent numerical modeling efforts led by \citet{2017ApJ...839...22H,2017A&A...601A.122D} and \citet{2019A&A...626A..33H} have confirmed that (QS)EBs are classic markers of small-scale magnetic reconnection events in the solar photosphere.   

In this study, we base our analysis on the small 100$\times$100 pixel FOV shown in panel~(a) of Fig.~\ref{fig:flux_emer_1} with a focus on the flux emergence event. Panel~(a) of Fig.~\ref{fig:QSEB1} shows the QSEB in the blue-wing \hbeta{} intensity map (indicated by the magenta marker). The animation shows a tiny, flame-like brightening lasting for about 40~s. We note that this period (along with the latter QSEB) is marked in panels~(e) and (f) of Fig.~\ref{fig:flux_emer_1} as QSEB~1 and 2 respectively. The \hbeta{} spectral-time ($\lambda-t$) slice in Fig.~\ref{fig:QSEB1}~(b) shows an enhancement in the line-wings compared to the background which is reflected in the spectra shown in panel~(c). The observed spectral shape is characteristic to an EB. Moreover, the co-temporal \hbeta{} WB image in panel~(d) shows no such brightening which clearly distinguishes this QSEB from a typical photospheric magnetic bright point. Panels~(e) and (f) show the location of the QSEB on the LOS magnetic field map and the evolution of the total positive and negative magnetic flux inside the smaller cyan box shown in panel~(e). From the animation and the light curves, we see that the negative flux decreases up to about 40~s from the start of the event after which it stabilizes up to $t=55$~s, following which it starts to decrease right around the onset of the QSEB. 

Figure~\ref{fig:QSEB2} shows another QSEB event (QSEB~2) associated with the same emerging flux region under consideration. QSEB~2 is brighter, but shorter-lived ($\approx$25~s) in comparison to QSEB~1 and it is shown against a background of red-wing \hbeta{} intensity map. Both the examples bear close morphological resemblance with distinct flame-like brightenings. The $\lambda-t$ slice and the spectra for QSEB~2 also show characteristic EB-like behavior but as is also evident from panel~(c), the intensity enhancement is stronger compared to QSEB~1. The corresponding WB image (panel~d) shows that QSEB~2 is located in the intergranular lane, and from the B$_{\mathrm{LOS}}$ map we find that in this case the QSEB exists in the intersection of opposite polarities. The variation of the negative polarity flux in panel~(f) shows an increase right around the onset of the QSEB.

The examples presented in this section clearly show that the emerging flux episode described in the previous section have a definite impact not just in the form of enhanced chromospheric spicular activity, but also deeper in the solar atmosphere where it reconnects and subsequently releases energy in the form of small-scaled EBs. 

\subsubsection{Flux emergence episode 2}
\label{subsubsection:FE_2}

\begin{figure*}[ht!]
% \plotone{Context_CBP_v2.png}
\includegraphics[width=\linewidth]{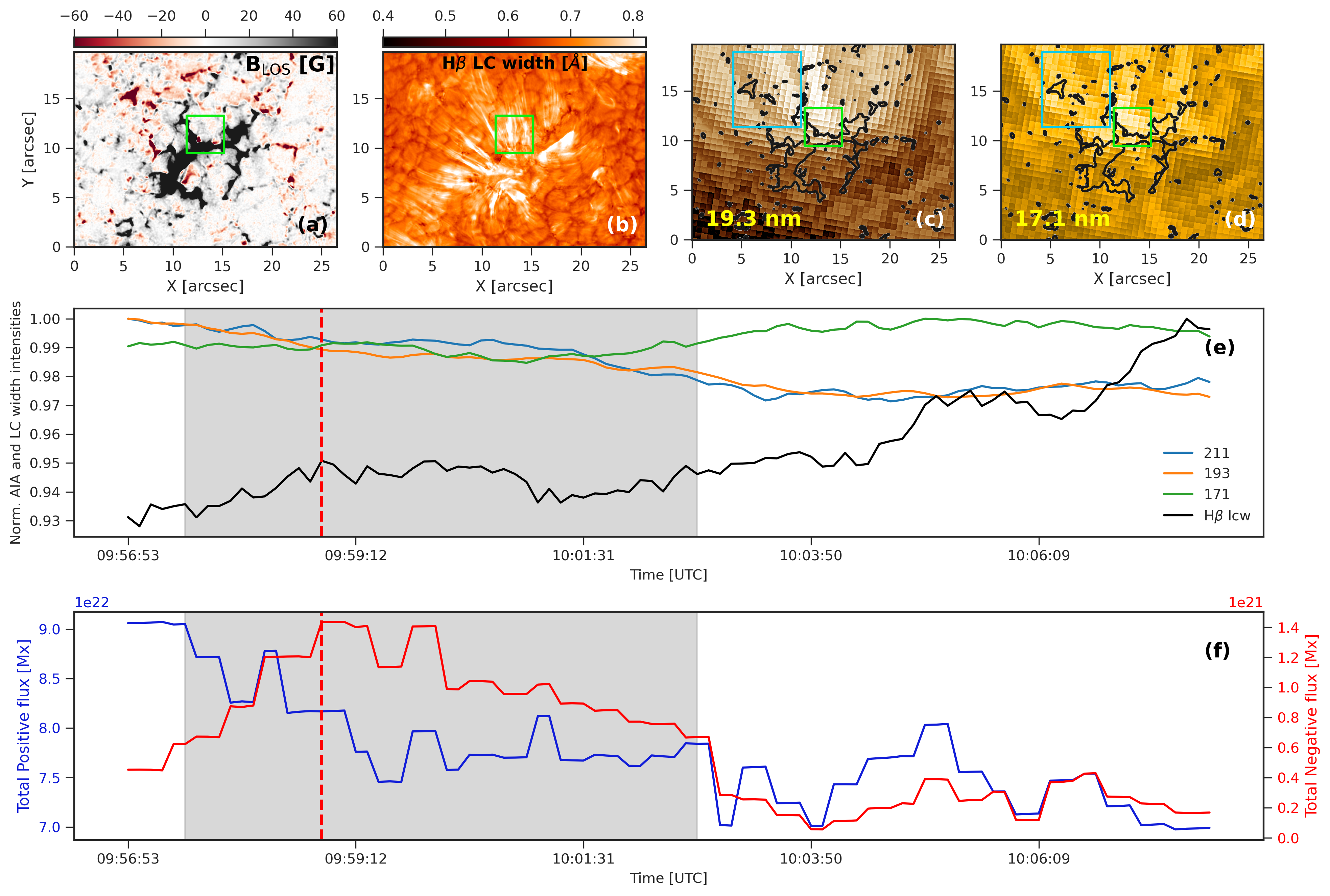}
\caption{Chromospheric and coronal response to flux emergence episode 2 in the same format as Fig.~\ref{fig:flux_emer_1}. No QSEBs were observed during this event. An animation of this figure is available \href{https://www.dropbox.com/s/ek7ul9twe7ts3re/FE_spicule_CBP_FOV3_new_emer.mp4?dl=0}{online}, which shows the flux emergence episode and corresponding chromospheric and coronal response for the entire 11~min duration in the same format as Fig.~\ref{fig:flux_emer_1}. \label{fig:flux_emer_2}}
\end{figure*}

In this section, we analyze the chromospheric and coronal responses to another flux emergence episode that occurred close to the footpoints of the CBP. Figure~\ref{fig:flux_emer_2} depicts the overview of the episode in the same format as Fig.~\ref{fig:flux_emer_1}, and the emergence is shown with a green colored box (occupying the same area as before) drawn in panel~(a). A close inspection of the animation linked with panel~(a) suggests a clear but relatively smaller negative flux emergence episode lasting $\approx$ 5.5~min. This is also evident from panel~(f) which shows an increase in the total negative flux within the cyan colored box starting around 09:57UT. The total negative flux increases by $\approx$180\% compared to the pre-event values and peaks around 09:59:12UT. It then starts to decrease steadily reaching a minimum value of 0.1$\times$10$^{21}$~Mx at 10:03:50UT. The emerging negative polarity, however, starts to disappear around 10:02:37UT (indicated by the gray region in panels~e and f) from the B$_{\mathrm{LOS}}$ map. 

We found a contrasting evolution of the light curves associated with the chromospheric and coronal channels for this emergence episode in comparison to the event described in Sect.~\ref{subsubsection:FE_1}. The \hbeta{} LC width intensity level in panel~(e) does not show a marked increase in tandem with the flux emergence and subsequent cancellation; it maintains a steady level during the whole flux emergence episode and only starts to increase well after the total negative flux reaches its minimum value. The dynamical evolution of the spicules seen in panel~(b) complements the variation of the \hbeta{} LC light curve indicated in panel~(e) where we do not see any significant enhancement in spicular activity compared to pre-emergence scenario. The coronal channels behave similarly where very little (or no) changes in their respective intensity levels are seen during the entire emergence event, implying that none of the channels are impacted directly by this emergence episode unlike the scenario outlined in Sect.~\ref{subsubsection:FE_1}. Again, to not be limited to a single FOV, we repeated the analysis by choosing different (rectangular) cyan FOVs in panels~(c) and (d) like before. However, we did not find any differences in the temporal variation of the AIA light curves. In addition, we were also not able to find any signature of QSEBs linked with this event. A possible explanation could be that the strength of the emerging flux in the second episode is at least a factor of three lesser than the first episode, which reinforces our conclusion that there is likely no impact on the chromosphere and the corona associated with this weaker flux emergence event. 

%It is not very straightforward to explain the cause of the observed differences between the responses to the two flux emergence events simply based on observations. A possible explanation could be that the strength of the emerging flux in the second episode is at least about a factor of three lesser than the first episode. Therefore, it is likely that such (weaker) emergence events have no direct relationship with the dynamics higher up in the solar atmosphere. Advanced high-resolution numerical simulations \citep[like][but in 3D]{2022ApJ...935L..21N} spanning from the photosphere to the corona are needed to have a clearer idea of the impact of such small-scaled emergence episodes. 

Identifying small-scale flux emergence events, such as the ones described in this paper, can be a challenging task. This is primarily because high-resolution observations of the solar photosphere reveal a myriad of magnetic features especially in the regions close to a network or an inter-network. This is somewhat clear from the LOS magnetic field maps used in this paper where we see sub-arcsecond fields appearing and disappearing all over the FOV. Therefore, it is imperative that future studies warrant the need for high-resolution telescopes (achieving accurate polarimetry at a resolution of 0\farcs2 or better) to discern the impact of such small-scale flux emergence episodes on the overlying coronal structures.

% Moreover, it is impossible to predict in advance (with sufficient accuracy) when and where an emerging flux event will occur. In this section, we presented two cases of flux emergence episodes by relying solely on visual inspection. The choice of the two regions was based on the close proximity of the emerging flux events to the dominant positive polarity -- which appears to be the root of the CBP structure.

% or appear in absorption due to the extinction of photons emitted from the hot coronal plasma in overlying \ion{H}{i}, \ion{He}{i}, and \ion{He}{ii} atoms in cooler plasma \citep[similar to the scenario shown in Fig.~10 of][]{1999ASPC..184..181R}. It is to be noted that \citet{2011ApJ...743...23M} found  

\section{Summary and conclusions}
\label{section:conclusions}

Despite several decades of scientific research dedicated to the study of CBP, their chromospheric counterparts remained largely unexplored with the exception of \citet{1981SoPh...69...77H}, and very recently \citet{2021A&A...646A.107M}. This paper is an attempt in that direction where the focus is on the chromosphere underneath a CBP observed at spatial and temporal scales that have never been reported before. In particular, this study primarily investigates the relationship between ubiquitous spicules seen at the footpoints of a CBP observed in the \hbeta{} spectral line, and their coronal counterparts. The chromospheric scenery reveals a conspicuous morphological and topological resemblance with the loops of the CBP, indicating that spicules form an integral part of the overall magnetic structure. This interpretation is further reinforced by computing 2D density distribution of over 6000 spicules detected using an automated procedure, and comparing them against coronal images. Our analysis reveals that these spicules predominantly lie close to the footpoints of the CBP and have the same orientation as their coronal counterparts.

We show illustrative examples indicating the ``connection" between spicules and CBP loops that is suggestive of the scenario that spicular flows are often associated with heating to TR and coronal temperatures, and can propagate into the corona likely in the form of PCDs \citep[][see Fig.~4]{2022ApJ...935L..21N}. Thus, they can potentially contribute toward transient intensity perturbations of an already existing CBP. Furthermore, we also show an example of a twist propagating in the CBP loops that is directly correlated with twisting spicules seen in the chromospheric footpoints. All such examples provide strong indication of a direct link that exists between the chromosphere and the corona of a CBP. It is however not straightforward to explain whether spicules observed at the footpoints of the CBP are unique compared to the spicules observed elsewhere. From past studies, it is expected that the strength of the magnetic field (and its inclination) in the lower atmosphere plays a major role in driving the observational properties of spicules. Statistical analysis by \citet{2012ApJ...759...18P} reveal clear differences between properties of spicules in active regions and quiet-Sun (and coronal holes), and \citet{2011ApJ...743..142H} report similar findings with numerical simulations. Underneath the CBP, the field strength is distinctly stronger compared the rest of the FOV--where the rapid expansion of the weaker field lines likely leads to different coronal impact. Statistical studies with coordinated chromospheric and higher-resolution coronal observations (e.g. from MUSE), in addition to detailed quantitative analysis of mass-energy exchanges, are needed to determine if the coronal contribution of spicules depend on the strength of the photospheric magnetic fields.

We also investigate the chromospheric and coronal responses to two different flux emergence episodes and find very different results. In the first case, we see a clear enhancement in the chromospheric spicular activity in tandem with the flux emergence event. The emission in the 17.1~nm channel shows a strong correlation with the chromospheric activity whereas the same cannot be said for the 19.3 and 21.1~nm channels. The emission in the latter two channels decreases (but only by about 3\%) almost co-temporally with the enhancement seen in the 17.1~nm channel. 
% This is attributed to the fact that the heating associated with the ensemble of chromospheric spicules observed during this event is likely not vigorous enough to contribute toward the emission from the dominant ions in the 19.3 and 21.1 passbands.
The second flux emergence episode does not seem to contribute toward either a change in the chromospheric or coronal activity. This is likely due to a weaker (and smaller-scaled) flux emergence compared to the previous episode which causes little to no impact in the upper atmospheres of the Sun. Further coordinated observations (along with numerical simulations) spanning the photosphere through the corona are needed to statistically establish as to when and why such small-scaled emergence episodes impact the CBP above.

We also found distinct signatures of magnetic reconnection associated with the stronger flux emergence episode in the form of multiple QSEBs. Although we found a slight co-temporal intensity increase in one of the coronal channels, it is not straightforward to correlate that directly with the reconnection happening in the upper photosphere. As explained before, the likely cause of such a coronal intensity enhancement is attributed to the enhanced chromospheric spicular activity seen in the chromosphere. 

% Flux emergence is often linked with heating in the coronal bright points. Such heating is attributed to magnetic reconnection in the coronal heights and is generally observed in the form of enhanced intensity emission of the coronal channels \citep[Sect.~9 of][]{2019LRSP...16....2M}. In this paper, however, we found signatures of magnetic reconnection relatively deeper in the solar atmosphere in the form of QSEBs linked to the stronger flux emergence episode. Though we found a slight co-temporal increase in one of the coronal channels, it is not straightforward to correlate that directly with the reconnection happening in the upper photosphere. As explained before, the likely cause of such a coronal intensity enhancement is attributed to the enhanced chromospheric spicular activity seen in the chromosphere. 

The results presented in this paper attempts to describe the (complex) chromospheric scenery underneath a CBP from the perspective of high-resolution observations for the very first time. However further studies, including both the footpoints of a CBP, are needed in coordination with ground- and space-based observations to answer some of the outstanding questions in more detail. ``Connecting" the photospheric magnetic footpoints to the corona through the chromosphere remains a challenge. Current instrumentation does not allow simultaneous photospheric, chromospheric, and coronal magnetic field measurements of sufficient spatial resolution and quality. Until that becomes feasible, a possible way forward is the use of non-potential magnetic field extrapolations in combination with 3D numerical simulations. Such comparisons may lead to a better understanding of how flux emergence impacts the chromosphere and the corona overlying a CBP.

\begin{acknowledgments}
S.B. and B.D.P. gratefully acknowledge support from NASA grant 80NSSC20K1272 "Flux emergence and the structure, dynamics, and energetics of the solar atmosphere".
We thank Edvarda Harnes for the SDO to SST alignment.
% SST
The Swedish 1-m Solar Telescope is operated on the island of La
Palma by the Institute for Solar Physics of Stockholm University in the Spanish
Observatorio del Roque de Los Muchachos of the Instituto de Astrofísica
de Canarias. The Institute for Solar Physics is supported by a grant for research
infrastructures of national importance from the Swedish Research Council (registration
number 2017-00625). 
% IRIS
% IRIS is a NASA small explorer mission developed
% and operated by LMSAL with mission operations executed at NASA Ames
% Research center and major contributions to downlink communications funded
% by ESA and the Norwegian Space Centre. 
% NFR : 250810 Toppforsk, 325491 ISSRESS, 262622 RoCS
This research is supported by the Research Council of Norway, project numbers 250810, 325491, and through its Centres of Excellence scheme, project number 262622.
D.N.S. acknowledges support by the European Research Council through the
Synergy Grant number 810218 (``The Whole Sun'', ERC-2018-SyG) and by
the Spanish Ministry of Science, Innovation and Universities through project
PGC2018-095832-B-I00. 
\end{acknowledgments}

\bibliography{sample631}{}
\bibliographystyle{aasjournal}



\end{document}